\documentclass[aps,prl,10pt,superscriptaddress,twocolumn,amsmath]{revtex4-2}

\usepackage[T1]{fontenc}

\usepackage{graphicx}
\usepackage{natbib}
\usepackage{siunitx}
\usepackage{bm}
\usepackage{enumitem}
\usepackage{textcomp}
\usepackage{amsmath} 
\usepackage{bbold} 

\setlist{nosep}
\usepackage[colorlinks=true,citecolor=blue,linkcolor=blue]{hyperref}

\newcommand{\ket}[1]{\ensuremath{\left| #1 \right\rangle}}
\newcommand{\NVz}[0]{NV$^{0}$}
\newcommand{\NVm}[0]{NV$^{-}$}

\begin{document}

\title{Low temperature photo-physics of single NV centers in diamond}
\author{Jodok Happacher}
\affiliation{Department of Physics, University of Basel, Klingelbergstrasse 82, Basel CH-4056, Switzerland}
\author{David Broadway}
\affiliation{Department of Physics, University of Basel, Klingelbergstrasse 82, Basel CH-4056, Switzerland}
\author{Patrick Reiser}
\affiliation{Department of Physics, University of Basel, Klingelbergstrasse 82, Basel CH-4056, Switzerland}
\author{Alejandro Jim\'enez}
\affiliation{Facultad de F\'isica, Pontificia Universidad Cat\'olica de Chile, Santiago 7820436, Chile}
\author{M\"arta A. Tschudin}
\affiliation{Department of Physics, University of Basel, Klingelbergstrasse 82, Basel CH-4056, Switzerland}
\author{Lucas Thiel}
\affiliation{Department of Physics, University of Basel, Klingelbergstrasse 82, Basel CH-4056, Switzerland}
\author{Dominik Rohner}
\affiliation{Department of Physics, University of Basel, Klingelbergstrasse 82, Basel CH-4056, Switzerland}
\author{Marcel.li Grimau Puigibert}
\affiliation{Department of Physics, University of Basel, Klingelbergstrasse 82, Basel CH-4056, Switzerland}
\author{Brendan Shields}
\affiliation{Department of Physics, University of Basel, Klingelbergstrasse 82, Basel CH-4056, Switzerland}
\author{Jeronimo R. Maze}
\affiliation{Facultad de F\'isica, Pontificia Universidad Cat\'olica de Chile, Santiago 7820436, Chile}
\author{Vincent Jacques}
\affiliation{Laboratoire Charles Coulomb, Université de Montpellier and CNRS, 34095 Montpellier, France}
\author{Patrick Maletinsky}
\email[]{patrick.maletinksy@unibas.ch}
\affiliation{Department of Physics, University of Basel, Klingelbergstrasse 82, Basel CH-4056, Switzerland}

\date{\today}

\begin{abstract}
	We investigate the magnetic field dependent photo-physics of individual Nitrogen-Vacancy (NV) color centers in diamond under cryogenic conditions.
	At distinct magnetic fields, we observe significant reductions in the NV photoluminescence rate, which indicate a marked decrease in the optical readout efficiency of the NV's ground state spin.
	We assign these dips to excited state level anti-crossings, which occur at magnetic fields that strongly depend on the effective, local strain environment of the NV center. 
	Our results offer new insights into the structure of the NVs' excited states and a new tool for their effective characterization. 
	Using this tool, we observe strong indications for strain-dependent variations of the NV's orbital $g$-factor, obtain new insights into  NV charge state dynamics, and draw important conclusions regarding the applicability of NV centers for low-temperature quantum sensing.
\end{abstract}

\maketitle

The nitrogen vacancy (NV) lattice defect in diamond\,\cite{Doherty2013a} hosts a versatile solid state spin system that finds applications in quantum metrology\,\cite{degenQuantumSensing2017}, nanoscale imaging\,\cite{Casola2018} or quantum information processing\,\cite{Kalb2017}. 
In these, the NV spin stands out due to its excellent quantum coherence properties, 
which persist across a wide range of temperatures\,\cite{Toyli2012a} and pressures\,\cite{Fu2020}. 
As a result of their performance and robustness, NV spins have been employed in practical applications ranging from remote spin-spin entanglement\,\cite{Hensen2015a} to nanoscale magnetic imaging\,\cite{Rondin2014,Casola2018}, even under cryogenic conditions\,\cite{Thiel2016a,Thiel2019a}.

The majority of such applications build on methods for efficient optical NV spin initialization\,\cite{Harrison2004, Tetienne2012} and readout\,\cite{Dreau2011,Steiner2010} -- two key features that result from properties of the NV's orbital excited states. 
At temperatures $T\gtrsim100~$K, orbital averaging allows for a description of the NV excited states as an effective spin-$1$ system, with spin states $\ket{-1}$, $\ket{0}$, $\ket{1}$, characterized by the magnetic quantum number along the Nitrogen-Vacancy axis\,\cite{Fu2009a,Batalov2009a}. 
Initialization and readout of the NV's ground-state spin-$1$ system then result from optical transitions being largely spin-conserving for $\ket{0}$, while excited state spin levels $\ket{\pm1}$ show a non-radiative intersystem crossing into NV spin singlet states, followed by relaxation into the NV triplet ground state\,\cite{Doherty2013a}.

At cryogenic temperatures, however, this effective spin-$1$ description of the NV excited states does not hold, since orbital averaging slows down and becomes negligible at temperatures $T\lesssim20~$K\,\cite{Batalov2009a,Fu2009a}. 
The question how the emerging and rich orbital excited state structure affects the NV's photo-physical properties is of relevance to most low-temperature experiments on NV spins, but has received remarkably little attention thus far. 
Previous work on NV ensembles has already attributed variations of NV photoluminescence (PL) with magnetic field to mixing of the excited states, but a complete picture was obscured by ensemble averaging\,\cite{Rogers2009a}. 
Conversely, for single NVs, only the regime near zero magnetic field and few fixed higher field values have
been explored thus far\,\cite{Batalov2009a,Pompili2021a}. 
At the same time, consistent observations of significant reductions in NV PL and spin-readout contrast have been made in the context of low-temperature NV magnetometry\,\cite{Thiel2019a,Lillie2020}, but remain unexplained up to now. 

\begin{figure*}[ht]
	\centering
	\includegraphics[width=\textwidth]{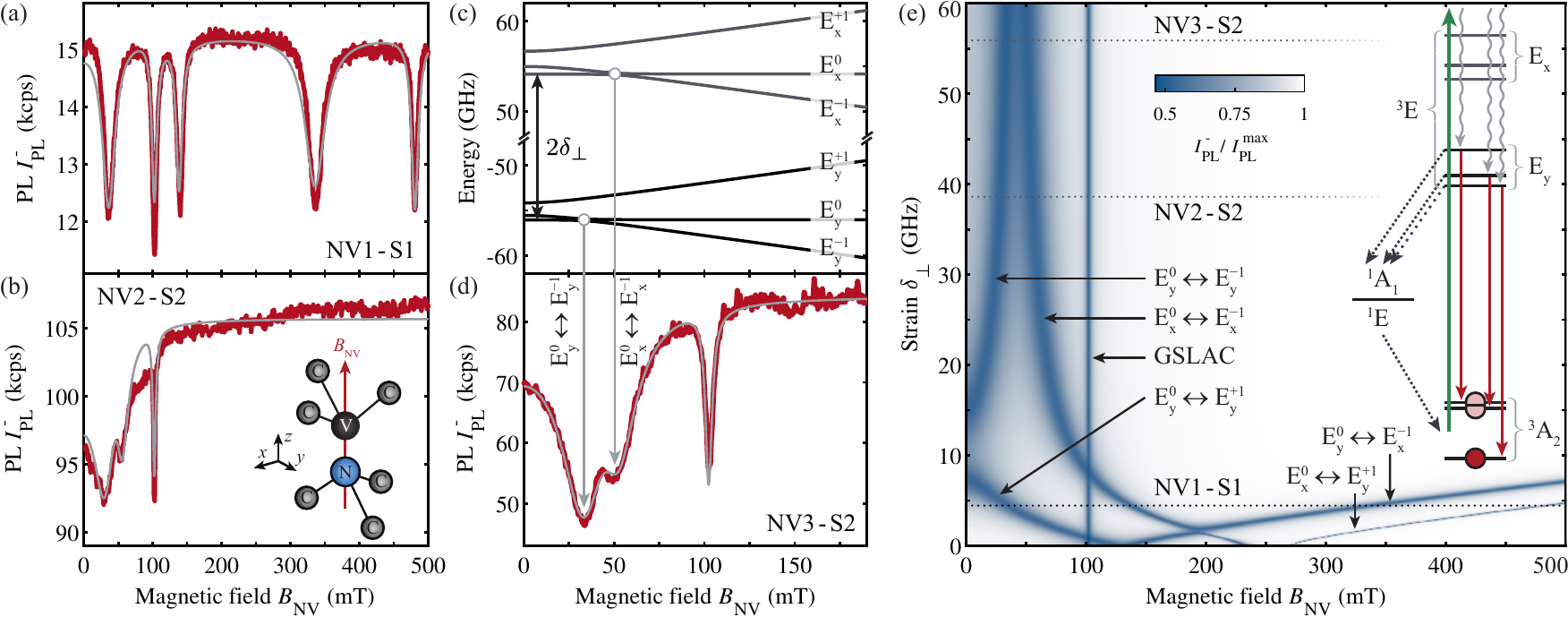}
	\caption{
		(a)~The \NVm{} photoluminescence (PL) signal, $I_{\rm PL}^{-}$, as a function of magnetic field, $B_{\rm NV}$, for a typical ``low strain'' NV, NV1-S1 (transverse strain parameter $\delta_\perp=4.444(3)~$GHz).
		(b)~Same as (a), but for an NV center exhibiting more strain ($\delta_\perp=38.6(7)~$GHz). 
		The common $I_{\rm PL}^{-}$ dip at $B_{\rm NV}=102.5~$mT in (a) and (b) originates from the well-known ground-state level anti-crossing (GSLAC), while additional dips result from excited-state level anti-crossings (ESLACs). Grey lines are fits to our model (see text and\,\cite{SOM}). All data were recorded at a temperature $T\approx 4~$K.
		(c)~Example of \NVm{} excited state energies for an NV experiencing even higher strain ($\delta_\perp=55.9~$GHz). Labels indicate the orbital- and spin degree of freedom of the states.
		(d)~$B_{\rm NV}$ dependence of $I_{\rm PL}^{-}$ for NV3-S2 with $\delta_\perp=55.9(6)~$GHz. The vertical arrows assign the observed $I_{\rm PL}^{-}$ dips to the corresponding ESLACs shown in (c).
		(e)~PL intensity $I_{\rm PL}^{-}$ as a function of $B_{\rm NV}$ and $\delta_\perp$, calculated from a classical rate equation model for the $10$-level system illustrated in (e) (for details see\,\cite{SOM}). 
		Labels in (e) indicate which ESLACs are responsible for the respective PL dips, while horizontal lines show the $\delta_\perp$-values for the NVs presented in panels (a), (b), and (d). 
		Inset:~Electronic levels and transition pathways considered in simulating the magnetic field dependence of $I_{\rm PL}^{-}$. 
		The \NVm{} excited state energies are illustrated for the limit of large strain. 
	}
	\label{fig:IvsB}
\end{figure*}

To address these questions, we present a systematic study of the photo-physical properties of individual NV centers 
at cryogenic temperatures.
Specifically, we study the dependence of the NV PL rate on static magnetic fields, $B_{\rm NV}$, applied along the NV axis.  
We observe significant PL reductions, at well-defined, strain-dependent values of $B_{\rm NV}$, that we assign to NV excited state level anti-crossings (ESLACs), which result in efficient NV spin mixing and subsequent intersystem crossings.
The $B_{\rm NV}$ values where the ESLACs occur further allow for an extraction of the NV's orbital $g$-factor, $g_l$.
Interestingly, we find previously reported values of $g_l$ to be inconsistent with our observations for NVs experiencing high strain, suggesting a strain dependence of $g_l$. 
In addition, our results provide new insights into (i) the mechanisms of NV charge state conversion and (ii) the efficiency of NV spin-initialization and readout, which are valuable for efficient NV-based quantum sensing under cryogenic conditions.

Our experiments are performed using a confocal optical microscope with samples held at temperatures $T\approx4$~K in a closed-cycle refrigerator with optical access and three-axis vector magnetic field control. 
Optical excitation is performed with green, 
continuous-wave laser excitation at power levels close to saturation of the NV's optical transition.
For optical detection, we use an avalanche photodiode and appropriate color filters to detect PL predominantly stemming from the NV's negative charge state, \NVm{}, with corresponding PL photon count rates $I_{\rm PL}^{-}$.

We first investigate individual NV centers in two $(100)-$oriented, single-crystal ``electronic grade'' diamonds (Element-Six), grown by chemical vapour deposition. 
These two samples differ in the nature of their NVs: 
In sample S$1$, we study naturally occurring NV centers, several microns deep in the bulk, while in sample S$2$, $\sim10~$nm deep NVs have been created by $^{14}$N$^+$ ion implantation at $12~$keV and subsequent annealing.
These two types of NV defects were chosen for studying a broad range of local strain, which is known to be increased for NVs implanted close to the diamond surface\,\cite{Broadway2018b}. 
To enhance NV PL collection efficiencies, diamond solid immersion lenses and nanopillars were structured on samples S$1$ and S$2$, respectively, with methods reported elsewhere\,\cite{Hedrich2020a,Robledo2011a,neuPhotonicNanostructures1112014}.

Figures\,\ref{fig:IvsB}a and b show representative data of $I_{\rm PL}^{-}$ as a function of $B_{\rm NV}$, for two NV centers, NV1-S1 and NV2-S2 
(where NV$k$-S$l$ denotes the $k$-th NV located in sample S$l$). 
For both NVs, $I_{\rm PL}^{-}(B_{\rm NV})$ shows several distinct, narrow local minima (``dips'') at specific values of $B_{\rm NV}$. 
The prominent, narrow dip at $B_{\rm NV}=102.5~$mT results from the NV's well-known ground-state level anti-crossing (GSLAC)\,\cite{Wickenbrock2016a} and is present for all NVs we investigated.
Conversely, the multiple additional dips, which occur reproducibly for each NV, only at different values of $B_{\rm NV}$, are thus far unaccounted for.

We attribute the observed $I_{\rm PL}^{-}$ dips (excluding the GSLAC) to 
ESLACs, which result in spin mixing and subsequent population shelving into the \NVm{} singlet manifold. 
Figures\,\ref{fig:IvsB}c,~d exemplify this concept for NV3-S2, where we show calculated \NVm{} excited state energies\,\cite{Batalov2009a,Doherty2013a,Maze2011a} (described below) alongside a measurement of $I_{\rm PL}^{-}(B_{\rm NV})$ to illustrate the coalescence between two ESLACs and the associated $I_{\rm PL}^{-}$ dips. 

To obtain further, quantitative insights, we employ an extended version of a classical rate-equation model of the NV's magnetic field-dependent photo-physics\,\cite{Tetienne2012,Rogers2009a}, where we now explicitly take into account the full, low-temperature excited state level structure of \NVm\,\cite{Maze2011a} (see Fig.\,\ref{fig:IvsB}e, inset, for the electronic states and population decay channels taken into account). 
For our calculations, we fix all transition rates to literature values by Gupta et al.\,\cite{guptaEfficientSignalProcessing2016}\footnote{We note that our main findings are insensitive to the rates employed and we find largely identical results by employing other literature values.}, and only leave the relevant parameters of the \NVm{} excited state Hamiltonian as free variables.

The energy splittings of the \NVm{} excited states sensitively depend on lattice strain transverse to the NV axis. Such strain can result from electric fields or crystal stress\,\cite{Maze2011a}, which we combine into a single perpendicular strain parameter $\delta_\perp=\sqrt{\delta_x^2+\delta_y^2}$, where $\delta_{x(y)}$ are the corresponding strain parameters along $x$ and $y$ (Fig.\,\ref{fig:IvsB}b, inset).
We then calculate the 
NV $^3E$ state energies by the Hamiltonian\,\cite{Doherty2013a}
\begin{equation}
\label{Eq:HES}
\hat{H}_{ES}=\hat{H}_{FS}+\delta_x\hat{\sigma}_z+\delta_y\hat{\sigma}_x+\mu_B(g_l\hat{\sigma}_y+g_e\hat{S}_z)B_{\rm NV},
\end{equation}
where $\hat{H}_{FS}$ is the fine structure Hamiltonian of the $^3E$ manifold\,\cite{Doherty2013a}, $\hat{\sigma}_i$ are the Pauli matrices representing the excited state orbital operators in the basis $\{E_x,E_y\}$ (the eigenstates of $\hat{H}_{ES}$ in the limit $\delta_\perp\gg||\hat{H}_{FS}||$), 
$\hat{S}_z$ is the $S=1$ spin operator along the $z$-axis, 
$\mu_B=28~$GHz/T is the Bohr magneton and $g_{l(e)}$ is the orbital (electron) $g$-factor ($g_e=2.01$\,\cite{Doherty2013a}).
Figure\,\ref{fig:IvsB}e shows the resulting model prediction of $I_{\rm PL}^{-}(B_{\rm NV})$ for varying $\delta_\perp$, and shows the strong dependence of the various $I_{\rm PL}^{-}$ dip locations on $\delta_\perp$. 
This explains the strongly varying $I_{\rm PL}^{-}(B_{\rm NV})$ traces we observe between different NVs, which naturally experience different levels of $\delta_\perp$. 
Conversely, a measurement of $I_{\rm PL}^{-}(B_{\rm NV})$ offers a sensitive tool to determine $\delta_\perp$ on the level of single NVs -- a task that otherwise requires complex spectroscopic techniques\,\cite{Batalov2009a}. 

We use our model to fit all $I_{\rm PL}^{-}(B_{\rm NV})$ data presented in this work, where grey lines 
overlaid on 
data show the resulting fits.
For the fits we only left $\delta_\perp$, and a small misalignment angle in $B_{\rm NV}$ (and $g_l$ in the case of NV1-S1 -- see below) as free parameters in Hamiltonian\,(\ref{Eq:HES}).
In addition, we allow for scaling constants for contrast and normalization of $I_{\rm PL}^{-}$ and a parameter describing the relative excitation efficiency into orbitals $E_x$ and $E_y$\,\cite{Fu2009a} to vary. 
All resulting fit parameters with errors are reported in\,\cite{SOM}. 
We note that despite these few degrees of freedom, our model yields excellent fits 
for NVs experiencing a large range of strain-values, from $\delta_\perp\lesssim ||\hat{H}_{FS}||$ (Fig.\,\ref{fig:IvsB}a) to $\delta_\perp\gg||\hat{H}_{FS}||$ (Fig.\,\ref{fig:IvsB}b).

\begin{figure}[ht]
	\centering
	\includegraphics[width=86mm]{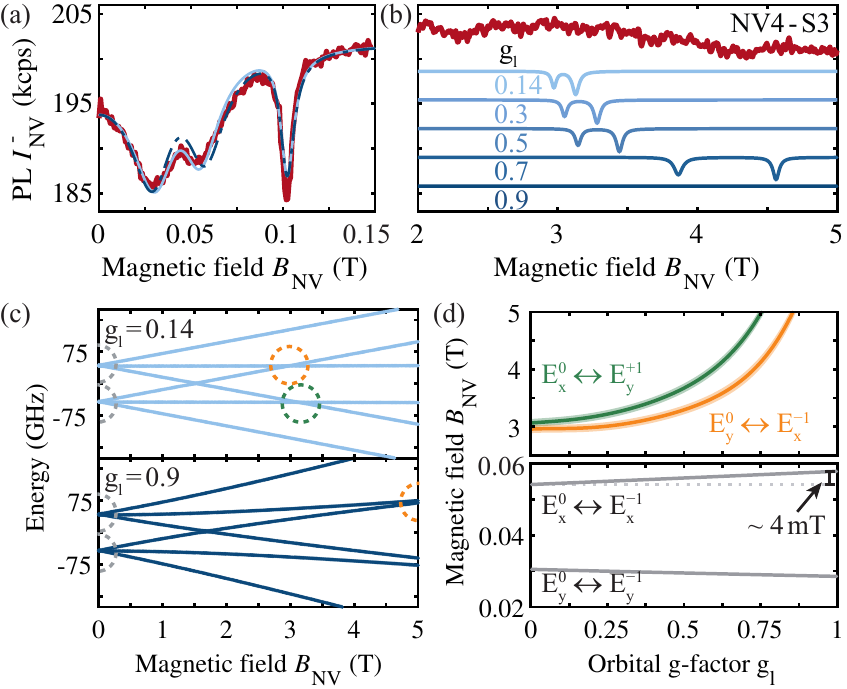}
	\caption{
		(a)~$I_{\rm PL}^{-}(B_{\rm NV})$ data (red) for NV4-S3 in the ``low-field'' regime ($B_{\rm NV}\lesssim150~$mT). The model fit (blue) yields $\delta_\perp=42.3(5)~$GHz and is largely unaffected by the value of $g_l$ ($g_l=0.14...0.9$; light to dark blue). 
		(b)~Same as (a) for stronger magnetic fields. 
		Blue-shaded curves show model predictions (vertically offset for clarity) for $\delta_\perp=42.3~$GHz and $g_l=0.14...0.9$ (see labels).
		(c)~Calculated $^3E$ excited state energies for $g_l=0.14$ and $g_l=0.9$. Dashed circles highlight the position of the $E_{x(y)}\leftrightarrow E_{y(x)}$ (green) and $E_{x(y)}\leftrightarrow E_{x(y)}$ (gray) ESLACs.
		(d)~$g_l$-dependence of the $B_{\rm NV}$ values at which various ESLACs occur. Shaded areas correspond to the propagated fitting errors in $\delta_\perp$ from (a).
	}
	\label{fig:HighField}
\end{figure}

However, we observe significant deviations when studying $I_{\rm PL}^{-}(B_{\rm NV})$ for more highly strained NVs at elevated magnetic fields.
There, our model predicts two additional $I_{\rm PL}^{-}$ dips (labelled $E_{x(y)}^0\leftrightarrow E_{y(x)}^{\pm1}$ in Fig.\,\ref{fig:IvsB}e), which arise from spin-mixing ESLACs between orbitals $E_x$ and $E_y$. 
To experimentally address this regime, we investigate single NV defects hosted in the $(111)-$oriented, electronic grade diamond S$3$\,\cite{Neu2014a}.
Figure\,\ref{fig:HighField}a shows low-field $I_{\rm PL}^{-}(B_{\rm NV})$ data for the representative NV4-S3, where a  fit (light blue) yields $\delta_\perp=42.3(5)~$GHz\,\cite{SOM},
for which our model predicts the $E_{x(y)}^0\leftrightarrow E_{y(x)}^{\pm1}$ dips to occur at Tesla-range magnetic fields (Fig.\,\ref{fig:HighField}b, light blue).
Importantly, the $E_{x(y)}^0\leftrightarrow E_{y(x)}^{\pm1}$ dip location depends sensitively on $g_l$ (blue traces in Fig.\,\ref{fig:HighField}b and Fig.\,\ref{fig:HighField}c), while the locations of the ``low-field'' $I_{\rm PL}^{-}$ dips shown in Fig.\,\ref{fig:HighField}a are largely independent of $g_l$ (Fig.\,\ref{fig:HighField}d).
Conversely, an observation of the $E_{x(y)}\leftrightarrow E_{y(x)}$ LACs yields a sensitive determination of $g_l$ on the level of single NVs. 
For NV4-S3 or any other NV we studied in sample S$3$, we were unable to observe the $E_{x(y)}^0\leftrightarrow E_{y(x)}^{\pm1}$ ESLACs, despite extensive experimental efforts; an observation which could be explained by an unexpected increase of $g_l$. 
Specifically, for the $\delta_\perp$ value of NV4-S3, $g_l\approx0.8$ is the lowest value consistent with our observations (Fig.\,\ref{fig:HighField}d), and we find similar conclusions for 
$\sim10$ further NVs with comparable values of $\delta_{\perp}$, where we examined $I_{\rm PL}^{-}(B_{\rm NV})$ for $B_{\rm NV}$ up to $5~$T\,\cite{SOM}. 

Our observations thus suggest a surprising, strain-induced enhancement of the orbital $g$-factor of \NVm{} over reported literature values $g_l\approx0.1...0.22$\,\cite{Reddy1987a, Rogers2009a, Braukmann2018a,Hanzawa1993a}.
Such an effect has not been discussed in literature thus far, but can be made plausible by qualitative arguments. 
At low strain, the NV's orbital excited states are near-degenerate, which allows for Jahn-Teller coupling to lead to a reduction of orbital angular momentum\,\cite{Braukmann2018a,Braukmann2018b}.
With increasing strain, the energy splitting of the states increases, which suppresses Jahn-Teller mixing, and thereby restores orbital angular momentum, leading to an increase of $g_l$ towards the classical value $g_l=1$.
While a complete theoretical description of this suggested $g_l$ enhancement is still lacking and beyond the scope of this work, we expect our findings to trigger further theoretical work on the topic.

\begin{figure}[ht]
	\centering
	\includegraphics[width=86mm]{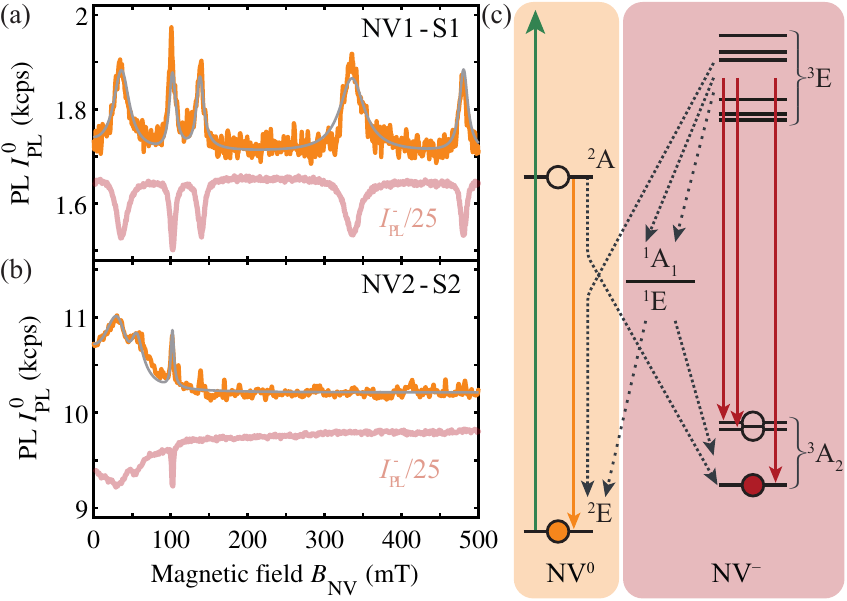}
	\caption{
		(a)~Magnetic field dependence of the \NVz{} PL rate, $I_{\rm PL}^{0}(B_{\rm NV})$, for NV1-S1 and 
		(b)~NV2-S2, along with the corresponding $I_{\rm PL}^{-}$-data (gray data-points taken from Fig.\,\ref{fig:IvsB}a,b, but rescaled for clarity). Gray lines are theory predictions based on the fits shown in Fig.\,\ref{fig:IvsB}a,b. 
		(c)~Electronic states of the \NVz{}-\NVm{} system, with corresponding (de-)ionization rates, used to include charge-state dynamics into the model introduced in Fig.\,\ref{fig:IvsB}f.
	}
	\label{fig:NV0}
\end{figure}

Our method of monitoring $I_{\rm PL}^{-}$ versus $B_{\rm NV}$ also offers insights into the charge dynamics of the NV center. 
Specifically, we conducted experiments where we chose appropriate color filters\,\cite{SOM} to record the PL 
intensity $I_{\rm PL}^0$ from the neutral charge state, \NVz{} (Fig.\,\ref{fig:NV0}a,b). 
Strikingly, we find that all previously described $I_{\rm PL}^{-}$ dips are mirrored by peaks in $I_{\rm PL}^0$. 
While the resulting buildup of \NVz{} population remains small (and therefore only minimally affects our conclusions thus far), our model can be extended to include previously reported \NVz{}$\leftrightarrow$\NVm{} (de)ionization processes\,\cite{Craik2020a,Aslam2013a} (Fig.\,\ref{fig:NV0}c). 
Importantly, within our model we can only explain the observed $I_{\rm PL}^0$ peaks by including a recently proposed\,\cite{Craik2020a}, decay channel from the \NVm{} singlet state $^1E$ to the ground state of \NVz{} -- without this process our model yields dips instead of peaks in $I_{\rm PL}^0$.
Using the previously measured (de)ionization rates\,\cite{Craik2020a,Aslam2013a}, our model then yields a quantitative  prediction for $I_{\rm PL}^0(B_{\rm NV})$, without any free fit parameters, other than scaling constants for contrast and  normalization of $I_{\rm PL}^0$ (grey lines in Fig.\,\ref{fig:NV0}a,b).

\begin{figure}[ht]
	\centering
	\includegraphics[width=86mm]{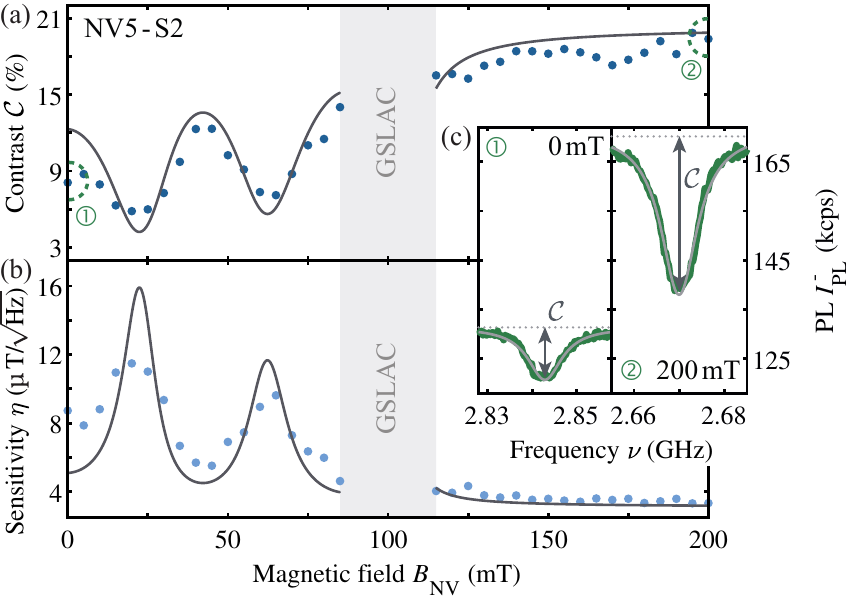}
	\caption{
		(a)~Dependence of the optically detected magnetic resonance (ODMR) contrast on magnetic field $B_{\rm NV}$ for NV5-S2. The microwave field amplitude (Rabi frequency) was held constant while varying $B_{\rm NV}$. 
		(b)~Magnetic field dependence of the magnetic field sensitivity extracted from data (see text). 
		Dark grey lines in a and b show theory predictions based on independently recorded $I_{\rm PL}^{-}(B_{\rm NV})$ data (see\,\cite{SOM}). 
		(c)~Exemplary ODMR traces recorded at values of $B_{\rm NV}$ (dotted circles in (a)), where magnetic field sensitivity is maximally (left) and minimally (right) affected by excited state level anticrossings. 
		}
	\label{fig:Magnetometry}
\end{figure}

The ESLAC induced dips in $I_{\rm PL}^{-}$ also have important implications for the magnetic-field sensitivity in low-temperature NV magnetometry.
Indeed, 
the dips in $I_{\rm PL}^{-}$ are accompanied by corresponding dips in the spin readout contrast $\mathcal{C}$ of optically detected magnetic resonance (ODMR)\,\cite{Gruber1997a}.
Figure\,\ref{fig:Magnetometry}a shows the evolution $\mathcal{C}(B_{\rm NV})$ for NV5-S2, while two exemplary ODMR traces recorded at and away from an $I_{\rm PL}^{-}$ dip ($B_{\rm NV}=0$ and $200~$mT, respectively) are shown in Fig.\,\ref{fig:Magnetometry}c.
This combined reduction in $I_{\rm PL}^{-}$ and $\mathcal{C}$  severely affect the NV's magnetic field sensitivity $\eta$.
Using the well-established estimate\,\cite{Dreau2011,Barry2019} $\eta = \frac{4}{3\sqrt{3}}\frac{\Delta\nu}{\gamma_{\rm NV} \mathcal{C} \sqrt{I_{\rm PL}^{-}}}$, where $\Delta\nu$ is the ODMR linewidth, and $\gamma_{\rm NV}=28~$GHz/T the NV's gyromagnetic ratio, we extract $\eta(B_{\rm NV})$ shown in Fig.\,\ref{fig:Magnetometry}b.
Compared to typical sensitivity values $\eta\approx3~\mu$T/Hz$^{0.5}$ away from the ESLACs (e.g. at $B_{\rm NV}\approx200~$mT), $\eta$ drops by almost an order of magnitude on the $I_{\rm PL}^{-}$ dips.

For nanoscale NV magnetometry, which exploits near-surface NV centers for which we consistently find $\delta_\perp>20~$GHz (Fig.\,\ref{fig:IvsB}f), this decrease of magnetometry performance predominantly affects magnetometry at $B_{\rm NV}\lesssim100~$mT (c.f. Fig.\,\ref{fig:IvsB}f). 
The fact that all such ``shallow'' NVs exhibit significant values of $\delta_\perp$ can be assigned to internal electric fields due to band bending near the diamond surface\,\cite{Broadway2018b,Stacey2018}, or to near-surface crystal stress\,\cite{McCloskey2020}. 
Magnetic field sensitivity for such NVs could in the future be restored by materials engineering to reduce $\delta_\perp$ near diamond surfaces, by working with NVs oriented normal to the diamond surface\,\cite{Rohner2019a} or by harnessing the dependence of the $I_{\rm PL}^{-}$ dip depth on the excitation laser polarization\,\cite{Fu2009a,SOM}.
Further improvements could be achieved by optimising laser excitation powers or by exploiting resonant laser excitation where specific ESLACs could be avoided by choosing the proper laser excitation frequency.

Our work on the low temperature magnetic field dependence of NV fluorescence rates offers a simple, yet precise and quantitative tool to characterize the excited state structure of individual NV centers.
These findings not only offer insights into the NV center orbital structure and charge dynamics, but are also relevant to applications in quantum information processing and quantum sensing, where precise knowledge of the excited state structure is key. 
Exploring the temperature-dependence of the $I_{\rm PL}^{-}$ dips would constitute a worthwhile extension to our work that might offer further insights into orbital averaging processes that dominate the NV's photo-physics at elevated temperatures\,\cite{Fu2009a}.
Importantly, the method we demonstrated and applied here is not limited to NV centers alone -- the excited state structure of any colour center exhibiting dark states that can be populated through magnetic field tunable ESLACs, such as the neutral Silicon-Vacancy center in diamond\,\cite{Zhang2020a}, could be investigated as well. 

We gratefully acknowledge Ronald Hanson, Arian Stolk and their colleagues for making sample S$1$ available for this study and Johannes K\"olbl for help in creating the figures. We further acknowledge financial support through the NCCR QSIT (Grant No. $185902$), the Swiss Nanoscience Institute, the EU Quantum Flagship Project ASTERIQS (Grant No. $820394$), and through the Swiss NSF Project Grant No. $188521$.

\bibliographystyle{apsrev4-1}
\bibliography{LT_photo-phys_Bibliography.bib}

\clearpage

\onecolumngrid
\appendix

\section*{Supplementary Information}

\tableofcontents

\section{Theoretical description}

\subsection{NV spin Hamiltonians}

To obtain further, quantitative insights, into the photo-physics of NV centers at low temperature we employ an extended version of a classical rate-equation model of the NV's magnetic-field dependent photo-physics\,\cite{Tetienne2012,Rogers2009a}, which explicitly takes into account the low-temperature excited state structure of \NVm{}\,\cite{Maze2011a}. 

For an appropriate description of the NV photo-dynamics at low temperatures one needs to consider both the ground and excited states Hamiltonians. 
Here we describe the model using the Hamiltonian previously described by Doherty et al\,\cite{Doherty2013a} but this process can also be performed using the alternative Hamiltonian form described by Maze et al\,\cite{Maze2011a}.

The canonical spin-Hamiltonian of the NV spin's ground state Hamiltonian is
\begin{equation}
\hat{\mathcal{H}}_{gs}=D_{gs}\Big[\hat{S}_{z}^{2}-S(S+1)/3~\mathbb{1}_{3}\Big] ,
\end{equation}
where $D_{gs} \approx 2.88~GHz$, $S = 1$ for a spin 1 system, $\hat{S}_{z}$ is the spin operator and $\mathbb{1}_{3}$ is the identity matrix.
The NV spin's internal hyperfine coupling and quadrupole moment are neglected as we observed no additional effect from these terms. 

The NV spin's ground state level structure is further modified by static electric ($\vec{E}$), magnetic ($\vec{B}$) and strain ($\vec{\delta}$) fields, whose contribution is given by
\begin{equation}
\begin{aligned}
\hat{\mathcal{V}}_{gs} = \: & {} \mu_{B} g^{\parallel }_{gs} \hat{S}_{z} B_{z}+
\mu_{B} g^{\bot }_{gs} \Big(\hat{S}_{x} B_{x}+\hat{S}_{y} B_{y}\Big)+
d^{\parallel }_{gs} (E_{z}+\delta_{z}) \Big[\hat{S}_{z}^{2}-S(S+1)/3~\mathbb{1}_{3}\Big]\\
& {} +d^{\bot }_{gs} (E_{x}+\delta_{x}) \Big(\hat{S}_{y}^{2}-\hat{S}_{x}^{2}\Big)+
d^{\bot }_{gs} (E_{y}+\delta_{y}) \Big(\hat{S}_{x} \hat{S}_{y}+\hat{S}_{y} \hat{S}_{x}\Big) ,
\end{aligned}
\end{equation}
where $\mu_{B}$ is the Bohr magnetron, $g^{\parallel }_{gs}$ and $g^{\bot }_{gs}$ are the components of the ground state electronic g-factor tensor, $d^{\parallel }_{gs}$ and $d^{\bot }_{gs}$ are the components of the ground state electric dipole moment.
The electric field and reduced stress tensor terms are treated as a single effective electric field.

The excited state of the nitrogen vacancy is an orbital doublet whose nature is masked at higher temperatures due to an orbital averaging.
The low temperature fine structure of the NV spin is given by the effective Hamiltonian,
\begin{equation}
\begin{aligned}
\hat{\mathcal{H}}_{es} = \: & {} \mathbb{1}_{2} \otimes D^{\parallel }_{es}\Big[\hat{S}_{z}^{2}-S(S+1)/3~\mathbb{1}_{3}\Big]\\
& {} -\lambda^{\parallel }_{es} \: \hat{\sigma}_{y} \otimes \hat{S}_{z}
+D^{\bot }_{es}\Big[ \hat{\sigma}_{z} \otimes \Big(\hat{S}_{y}^{2}-\hat{S}_{x}^{2}\Big) - \hat{\sigma}_{x} \otimes \Big(\hat{S}_{x} \hat{S}_{y}+\hat{S}_{y} \hat{S}_{x}\Big) \Big]\\
& {} +\lambda^{\bot }_{es}\Big[ \hat{\sigma}_{z} \otimes \Big(\hat{S}_{x} \hat{S}_{z}+\hat{S}_{z} \hat{S}_{x}\Big) - \hat{\sigma}_{x} \otimes \Big(\hat{S}_{y} \hat{S}_{z}+\hat{S}_{z} \hat{S}_{y}\Big) \Big] ,
\end{aligned}
\end{equation}
where $\sigma_{x,y,z}$ is the standard two level Pauli spin matrices, $D_{es}^{\parallel }$ and $D_{es}^{\bot }$ are the spin-spin interaction terms, and  $\lambda^{\parallel }_{es}$ and $\lambda^{\bot }_{es}$ are spin-orbit interaction terms.  
This excited state Hamiltonian results in two spin-1 systems, one for each of the two orbital branches.

The influence of external field on the NV spin's excited states level structure is given by,
\begin{equation}
\begin{aligned}
\hat{\mathcal{V}}_{es}^{LT} = \: & {} d^{\parallel }_{es}\Big(E_{z}+\delta_{z}\Big)~\mathbb{1}_{2} \otimes \mathbb{1}_{3}
+d^{\bot }_{es}\Big(E_{x}+\delta_{x}\Big)~\hat{\sigma}_{z} \otimes \mathbb{1}_{3}
-d^{\bot }_{es}\Big(E_{y}+\delta_{y}\Big)~\sigma_{x} \otimes \mathbb{1}_{3}\\
& {} +\mu_{B} \: l^{\parallel }_{es} \: B_{z} \: \hat{\sigma}_{y} \otimes \mathbb{1}_{3}
+\mathbb{1}_{2} \otimes \mu_{B} \: g^{\parallel }_{es} \: B_{z}~\hat{S}_{z}
+\mu_{B}\: g^{\bot }_{es} \Big( B_{x} \hat{S}_{z} +  B_{y} \hat{S}_{y}\Big)
\end{aligned}
\end{equation}
where $d^{\parallel }_{es}$ and $d^{\bot }_{es}$ are components of the electronic dipole moment, $l^{\parallel }_{es}$ is the orbital magnetic moment also referred as $g_l$, the effective orbital g-factor and $g^{\parallel }_{es}$ and $g^{\bot }_{es}$ are components of the electronic g-factor tensor.

\subsection{Transition rate equations}\label{Section: rates}


For an appropriate description of the photo-dynamics of a single NV at low temperatures one needs to consider the interplay between the three levels in the ground state, the six levels in the two orbital branches of the excited state and the metastable singlet.
The singlet can be reduced to one state with rates in and out of it\,\cite{Goldman2015b}, leaving ten states. 
We employ an extended version of a classical rate-equation model of the NV's magnetic-field dependent photo-physics\,\cite{Tetienne2012,Rogers2009a}.

The states for the rate equation are defined as:\\
Ground state:
\begin{align*}
	\ket{1} &\equiv {^3}A_2^0 \\
	\ket{2} &\equiv {^3}A_2^{-1} \\
	\ket{3} &\equiv {^3}A_2^{+1}
\end{align*}
Excited state:
\begin{align*}
	\ket{4} &\equiv {^3}E_y^0 \\
	\ket{5} &\equiv {^3}E_y^{-1} \\
	\ket{6} &\equiv {^3}E_y^{+1} \\
	\ket{7} &\equiv {^3}E_x^{0} \\
	\ket{8} &\equiv {^3}E_x^{-1} \\
	\ket{9} &\equiv {^3}E_x^{+1}
\end{align*}
Combined singlet state:
\begin{align*}
	\ket{10} &\equiv {^1}A_1/{^1}E \\
\end{align*}
Where the Eigenstates are give in the same notation as the main text. 
Which is $X^{m_s}$, where $X$ including the pre superscript and post subscript denote the orbital branch and the post subscript $m_s$ denotes the spin level, e.g. $E_y^{-1} \equiv \ket{E_y, -1} = \ket{\text{orbital state}, \text{spin state}}$.

The NV$^{-}$ can be optically excited with a green laser through a dipole-allowed transition to the excited state, which is a spin conserving.
This excitation does have a polarization-dependence, which is due to selection rules.
However, this dependence is suppressed due to the non-radiative decay observed from the non-resonant optical excitation\,\cite{Fu2009a}, leading to a small effect on the populations in the two excited state orbital-branches. 
Once in the excited state manifold, the NV spin can decay back to the ground state through two different decay channel: a radiative and an non-radiative decay.
The radiative decay is directly from the excited state back to the ground state.
While the non-radiative decay path is through the inter-system crossing (ISC) to a metastable singlet state, where a difference in the rates from the $m_s =0$ versus $m_s = \pm 1$ spin states is observed\,\cite{Tetienne2012}.


We begin with transition rates between ground- and excited-states with zero external field, i.e. where the spin quantum number is still well defined.
Where these rates have been experientially measured\,\cite{Robledo2011a,Tetienne2012,guptaEfficientSignalProcessing2016}.
The basis of the transition matrix are the ten eigenvectors $\ket{i^{0}}$ with $i = 1,...,10$. labeled with subscript $^0$ indicating that this is the basis at zero field.

In the following we make the assumption that spin conserving relaxation rates from the excited to ground states are the same for each spin state and that the non-spin conserving transitions are zero.
The non-spin conserving transitions rates have previously been shown to only be a few percent compared to the spin-conserving one\,\cite{Robledo2011a}.

The zero-field pumping rate can be assumed to be the respective relaxation rates from the excited state to the ground state $k_{r}$ (where the $r$ stands for radiative) multiplied by a pumping parameter $\beta$ which is proportional to laser power.
To capture the polarization dependence of the excitation we separate this pumping parameter into one for each orbital state $\beta_{E_{x}}$ and $\beta_{E_{y}}$.
The transition rates from the ground to excited states are thus defined as

\begin{equation}	
\begin{aligned}[c]
k^{0}_{\ket{1^{0}} \to \ket{4^{0}}} &=\beta_{E_{y}} \: k_{r}\\
k^{0}_{\ket{2^{0}} \to \ket{5^{0}}} &=\beta_{E_{y}} \: k_{r}\\
k^{0}_{\ket{3^{0}} \to \ket{6^{0}}} &=\beta_{E_{y}} \: k_{r}
\end{aligned}
\qquad\qquad\qquad
\begin{aligned}[c]
k^{0}_{\ket{1^{0}} \to \ket{7^{0}}} &=\beta_{E_{x}} \: k_{r}\\
k^{0}_{\ket{2^{0}} \to \ket{8^{0}}} &=\beta_{E_{x}} \: k_{r}\\
k^{0}_{\ket{3^{0}} \to \ket{9^{0}}} &=\beta_{E_{x}} \: k_{r}
\end{aligned}
\end{equation}

where the direct radiative decay transition rates from the excited to ground states are define as
\begin{equation}
\begin{aligned}[c]
k^{0}_{\ket{4^{0}} \to \ket{1^{0}}} &=k_{r}\\
k^{0}_{\ket{5^{0}} \to \ket{2^{0}}} &=k_{r}\\
k^{0}_{\ket{6^{0}} \to \ket{3^{0}}} &=k_{r}
\end{aligned}
\qquad\qquad\qquad
\begin{aligned}[c]
k^{0}_{\ket{7^{0}} \to \ket{1^{0}}} &=k_{r}\\
k^{0}_{\ket{8^{0}} \to \ket{2^{0}}} &=k_{r}\\
k^{0}_{\ket{9^{0}} \to \ket{3^{0}}} &=k_{r}
\end{aligned}
\end{equation}

The transition rates from the excited states to the metastable state are spin dependent and defined as
\begin{equation}
\begin{aligned}[c]
k^{0}_{\ket{4^{0}} \to \ket{10^{0}}} &=k_{nr_{0}}\\
k^{0}_{\ket{5^{0}} \to \ket{10^{0}}} &=k_{nr_{\pm 1}}\\
k^{0}_{\ket{6^{0}} \to \ket{10^{0}}} &=k_{nr_{\pm 1}}
\end{aligned}
\qquad\qquad\qquad
\begin{aligned}[c]
k^{0}_{\ket{7^{0}} \to \ket{10^{0}}} &=k_{nr_{0}}\\
k^{0}_{\ket{8^{0}} \to \ket{10^{0}}} &=k_{nr_{\pm 1}}\\
k^{0}_{\ket{9^{0}} \to \ket{10^{0}}} &=k_{nr_{\pm 1}}
\end{aligned}
\end{equation}
where $ k_{nr_{0}} \ll k_{nr_{\pm 1}}$\,\cite{Tetienne2012}.

The rates from the metastable state to the ground states are similar for all spin states and defined as,
\begin{equation}
\begin{aligned}[c]
k^{0}_{\ket{10^{0}} \to \ket{1^{0}}} &=k_{m_{0}}\\
k^{0}_{\ket{10^{0}} \to \ket{2^{0}}} &=k_{m_{\pm 1}}\\
k^{0}_{\ket{10^{0}} \to \ket{3^{0}}} &=k_{m_{\pm 1}}
\end{aligned}
\end{equation}
where $k_{m_{0}} \sim k_{m_{\pm 1}}$.

In a typical optically detected magnetic resonance (ODMR) measurement the spin population is transferred between the spin states of the ground state via an applied microwave field, and are defined as
\begin{equation}
\begin{aligned}[c]
k^{0}_{\ket{1^{0}} \to \ket{2}^{0}} &=k^{0}_{\ket{2^{0}} \to \ket{1^{0}}}=k_{MW_{-1}}\\
k^{0}_{\ket{1^{0}} \to \ket{3}^{0}} &=k^{0}_{\ket{3^{0}} \to \ket{1^{0}}}=k_{MW_{1}}
\end{aligned}
\end{equation}
where $k_{MW_x}$ is the driving transition rate on resonance with the $x$ transition between the $\ket{0}$ and  $\ket{x}$ states.
These rates are only non-zero when modeling the effect of the level anti crossings on ODMR in the main text. 

\begin{table}\label{Table:Rates}
	\begin{tabular}{l  l  l  l  l  l}
		\hline
		Reference &  $k_{r}$ \quad\quad & $k_{nr_{0}}$ \quad\quad & $k_{nr_{1}}$ \quad\quad & $k_{m_{0}}$ \quad\quad &  $k_{m_{\pm 1}}$ \quad\\
		\hline 
		 Robledo et al.\,\cite{robledoSpinDynamicsOptical2011} \quad\quad & 65 & 11 & 80 & 3.0 & 2.6 \\
		 Tetienne et al.\,\cite{Tetienne2012} \quad\quad & 65.9 & 7.9 & 53.3 & 0.98 & 0.73 \\
		 Gupta et al.\,\cite{guptaEfficientSignalProcessing2016} \quad\quad & 66.8 & 10.5 &90.7 &4.8 &2.2\\
		 \hline
	\end{tabular}
\caption{Experimentally measured transition rates at zero field. All of the rates are in MHz.}
\end{table}

All these transition rates have been measured experimentally, shown in Table\,\ref{Table:Rates}. 
In the model we used the parameters from Gupta et al.\,\cite{guptaEfficientSignalProcessing2016}, which we found the best agreement with our data.

We can now define the transition rate matrix at zero-field in this 10-level model.
The transition matrix, $Q^{0}$, gives the rates at zero electric, magnetic and strain field and is defined as
\begin{equation}
Q^{0}=
\begin{pmatrix}
0 & k_{MW_{-1}} & k_{MW_{1}} & \beta_{E_{x}} \; k_{r} & 0 & 0 & \beta_{E_{y}} \; k_{r} & 0 & 0 & 0 \\ 
k_{MW_{-1}} & 0 & 0 & 0 & \beta_{E_{x}} \; k_{r} & 0 & 0 & \beta_{E_{y}} \; k_{r} & 0 & 0 \\ 
k_{MW_{1}} & 0 & 0 & 0 & 0 & \beta_{E_{x}} \; k_{r} & 0 & 0 & \beta_{E_{y}} \; k_{r} & 0 \\ 
k_{r} & 0 & 0 & 0 & 0 & 0 & 0 & 0 & 0 & k_{nr_{0}} \\ 
0 & k_{r} & 0 & 0 & 0 & 0 & 0 & 0 & 0 & k_{nr_{\pm 1}} \\ 
0 & 0 & k_{r} & 0 & 0 & 0 & 0 & 0 & 0 & k_{nr_{\pm 1}} \\ 
k_{r} & 0 & 0 & 0 & 0 & 0 & 0 & 0 & 0 & k_{nr_{0}} \\ 
0 & k_{r} & 0 & 0 & 0 & 0 & 0 & 0 & 0 & k_{nr_{\pm 1}} \\ 
0 & 0 & k_{r} & 0 & 0 & 0 & 0 & 0 & 0 & k_{nr_{\pm 1}} \\ 
k_{m_{0}} & k_{m_{\pm 1}} & k_{m_{\pm 1}} & 0 & 0 & 0 & 0 & 0 & 0 & 0 
\end{pmatrix}.
\end{equation}

This zero field transition matrix is extend to arbitrary fields by calculating the spin-state overlap. 
Since the new eigenstates are a linear combination of the previous eigenstates the scalar product is used to evaluate the corresponding overlap,
\begin{equation}
\ket{i}=\sum_{j=1}^{10}\alpha_{ij}\ket{j^{0}}
\end{equation}
where $\alpha_{ij}$ is the projection of the new state $\ket{i}$ onto the zero-field basis states $\ket{j}$.
This resulting transition matrix which depends on the magnetic field $\bm{B}$, the electric field $\bm{E}$ and the stress field $\bm{\delta}$ is:
\begin{equation}
Q(\bm{B},\bm{E},\bm{\delta})_{ij}=\sum_{k=1}^{10}\sum_{l=1}^{10}\alpha_{ik}^{2}Q^0_{kl}\alpha_{jl}^{2}.
\end{equation}

With this transition matrix a classical rate equation for the system can be defined where $N_{i}$ is the population in each state and $\bm{M}$ is the rate equation matrix. We assume that this is a closed loop system, which impose the inclusion of diagonal terms such that,
\begin{equation}
M_{ij}=
  \begin{cases}
  \sum_{k=1}^{10}-Q_{ik} & \text{if } i = j\\
  Q_{ji} & \text{if } i \ne j
  \end{cases}
\end{equation}

The rate equation problem is stated in the following equation:
\begin{equation}
\frac{dN_{i}}{dt}=\bm{M} N_{i}
\end{equation}
Where the populations in the steady state is given by the smallest eigenvalue and is normalised such that
\begin{equation}
\hat{N}_{i}=\frac{N_{i}}{\sum_{j=1}^{10} N_j}.
\end{equation}

The photoluminescence of the \NVm{}, $I_{\rm PL}^-$, is calculated by summing over the relevant radiative transitions from the excited state to the ground state and populations of these states, such that 
\begin{equation}
I_{\rm PL}^-=\sum_{i=4,5,6}\sum_{j=1}^{3}Q_{ij}\hat{N}_{i}+\sum_{i=7,8,9}\sum_{j=1}^{3}Q_{ij}\hat{N}_{i}.
\end{equation}
Additional modification to the photoluminescence such as background fluorescence, $I_{\rm bck}$, and collection efficiency, $\eta_{\rm collection}$, can also be introduced, such that 
\begin{equation}
	I^{-}_{\rm PL} \equiv I^{-}_{\rm Total} = \eta_{\rm collection} I_{\rm PL}^- + I_{\rm bck}.
\end{equation}

The ODMR contrast $\mathcal{C}$ is calculated by the photoluminescence $I^{-}_PL$ with and without a driving field in the grounds state. For example, with a microwave driving between $\ket{1}$ (${^3}A_2^0$) and $\ket{2}$ (${^3}A_2^{-1}$), the contrast can be determined by calculating,
\begin{equation}
\begin{aligned}[c]
\mathcal{C} =\frac{I^{-}_{\rm PL}(k_{MW_{-1}}=0) -I^{-}_{\rm PL}(k_{MW_{-1}}\neq0)}{I_{\rm PL}(k_{MW_{-1}}=0)}
\end{aligned}
\end{equation}

\subsection{Extension of rate equation model to include NV$^0$}\label{Sec: NV0}


To account for possible charge state change between  \NVm{} and \NVz{} we extend the model to include the energy levels of the \NVz{} as well as transition rates between the two charge states.
The \NVz{}  consists of two eigenstates, one ground state $\ket{11}$ and an excited state $\ket{12}$:
\begin{align*}
	\ket{11} &\equiv {^2}E\\
	\ket{12} &\equiv {^2}A
\end{align*}
The transition within the \NVz{} charge state and between the two charge states are defined as follows:
  
From the excited state of the \NVz{} there is a direct radiative decay down to the \NVz{} ground state:
\begin{equation} 
\begin{aligned}[c]
k_{\ket{12^{0}} \to \ket{11^{0}}} &=k_{r_{NV^{0}}}
\end{aligned}
\end{equation}
The transition rate from the ground state to the excited state of the \NVz{} is defined as the product of the radiative transition rate and the pumping power $\beta_{NV^0}$, such that
\begin{equation} 
\begin{aligned}[c]
k_{\ket{11^{0}} \to \ket{12^{0}}} &=\beta \: k_{r_{NV^{0}}}.
\end{aligned}
\end{equation}
The recombination rate from the \NVz{} excited state to the \NVm{} ground state is given by
\begin{equation} 
\begin{aligned}[c]
k_{\ket{12^{0}} \to \ket{1^{0}}} &=k_{\rm recomb}\\
k_{\ket{12^{0}} \to \ket{2^{0}}} &=k_{\rm recomb}\\
k_{\ket{12^{0}} \to \ket{3^{0}}} &=k_{\rm recomb}
\end{aligned}
\end{equation}
The ionization rate from the \NVm{} excited state to the \NVz{} ground state is given by
\begin{equation}
\begin{aligned}[c]
k_{\ket{4^{0}} \to \ket{11^{0}}} &=k_{\rm ion}\\
k_{\ket{5^{0}} \to \ket{11^{0}}} &=k_{\rm ion}\\
k_{\ket{6^{0}} \to \ket{11^{0}}} &=k_{\rm ion}
\end{aligned}
\qquad\qquad\qquad
\begin{aligned}[c]
k_{\ket{7^{0}} \to \ket{11^{0}}} &=k_{\rm ion}\\
k_{\ket{8^{0}} \to \ket{11^{0}}} &=k_{\rm ion}\\
k_{\ket{9^{0}} \to \ket{11^{0}}} &=k_{\rm ion}
\end{aligned}
\end{equation}
The recombination rate and ionization rate in the dark between the ground state of the \NVz{} and the ground state of the \NVm{}. 
\begin{equation} 
\begin{aligned}[c]
k_{\ket{1^{0}} \to \ket{11^{0}}} &=k_{\rm darkion}\\
k_{\ket{2^{0}} \to \ket{11^{0}}} &=k_{\rm darkion}\\
k_{\ket{3^{0}} \to \ket{11^{0}}} &=k_{\rm darkion}
\end{aligned}
\qquad\qquad\qquad
\begin{aligned}[c]
k_{\ket{11^{0}} \to \ket{1^{0}}} &=k_{\rm darkrecomb}\\
k_{\ket{11^{0}} \to \ket{2^{0}}} &=k_{\rm darkrecomb}\\
k_{\ket{11^{0}} \to \ket{3^{0}}} &=k_{\rm darkrecomb}
\end{aligned}
\end{equation}
where for continuous wave excitation the dark ionization and recombination rates are effective zero, $k_{\rm darkrecomb} = k_{\rm darkion} = 0$.

Additionally, we include a newly postulated shelf ionization transition\,\cite{Craik2020a} which goes from the metastable state to the \NVz{} ground state and is laser driven:
\begin{equation}
\begin{aligned}[c]
k_{\ket{10^{0}} \to \ket{11^{0}}} &=k_{\rm shelfion}
\end{aligned}
\end{equation}

The rates for these charges state from the literature adapted to our nomenclature are found in Table\,\ref{Table: charge state Rates}.

\begin{table}\label{Table: charge state Rates}
	\begin{tabular}{l  l  l  l  l  l l}
		\hline
		Reference & $k_{r_{NV^{0}}}$ \quad\quad & $k_{\rm ion}$ \quad\quad & $k_{\rm recomb}$ \quad\quad & $k_{\rm shelfion}$ \quad\quad &  $k_{\rm darkion}$ \quad\quad & $k_{\rm darkrecomb}$\\
		\hline 
		Craik, et al.\,\cite{Craik2020a} \quad\quad & 50 & $0.037 \beta k_{r}$ \quad\quad & $0.8\beta k_{r}$ \quad\quad & $0.36 \beta k_{r}$  \quad\quad& 100 & 300 \\
		\hline
	\end{tabular}
	\caption{Experimentally measured transition rates for the \NVz{} and the transitions between \NVz{} and \NVm{}. All of the rates are in MHz.}
\end{table}

The photoluminescence, $I_{\rm PL}^0$, is calculated by looking at the population and the transition rate of the radiative transition of the \NVz{}.
The rates are given by the corresponding transition matrix $Q(\bm{B},\bm{E},\bm{\delta})_{ij}$.
\begin{equation}
I^{0}_{\rm PL}=Q_{11,\;12}\;\hat{N}_{12}
\end{equation}
Where both background counts and collection efficiency can be included in the same fashion as in the \NVm.

\newpage
\section{Description of fitting methods}

\subsection{Magnetic field dependent photoluminescence of the NV$^-$}\label{Sec: FittingNVm}

Using the model described in Section\,\ref{Section: rates}, we fit the PL spectra as a function of magnetic field, using the rates from Ref.\,\cite{guptaEfficientSignalProcessing2016}. 
In general, the model is over constrained and as a consequence requires some approximations in order to fit accurately and physically.
The full model would include 14 parameters: three magnetic field components, three electric field components, three stress parameters (shear stress is ignored), two pumping rates, the collection efficiency, the background counts, and the effective g-factor.
We treat the combination of the electric field and the stress as one effective stress field, reducing the six corresponding parameters to three.
Furthermore, the positions of the ESLACs are independent of the $z$-component of the effective strain parameter, which is thus disregarded from the fit. 
We parametrize the magnetic field in spherical coordinates, with the tilt angle of the magnetic field away from the NV quantization axis denoted as $\theta_B$. In our experiments, the magnetic field alignment is such that $\theta_B \sim 0$, in which case the azimuthal angle $\phi_B$ of the magnetic field has no significant effect on the NV excited state spectrum and we set $\phi_B = 0$.
Finally, we define the photon collection efficiency $\eta$ to be constant and identical to all NVs and set $\eta=0.005$. 
This leaves the fit with only 8 parameters, such that:
\begin{equation}
	I^{-}_{\rm PL} = f(B_{NV}, \theta_B, \delta_\perp, \phi_\delta, \beta_{E_{x}}, \beta_{E_{y}}, I_{\rm bck}, g_l )
\end{equation}
where $\delta_\perp = \sqrt{\delta_x^2 + \delta_y^2}$ and $\phi_\delta$ is an angle such that $\delta_x=\delta_\perp\sin(\phi_\delta)$ and $\delta_y=\delta_\perp\cos(\phi_\delta)$. 

As discussed in the main text, the effective g-factor, $g_l$, can only be confidently determined in the low stress regime. We thus let $g_l$ vary as a free fit parameter for NV1-S1 only, but keep this parameter fixes for all NVs with more elevated strain values. There, we set $g_l=0.14$ -- the value determined on NV1-S1. 
As we argue in the main text, $g_l$ varies with strain, and fixing $g_l$ for these NVs might therefore introduce certain systematic errors in our fits. However, as illustrated in Fig.2d (lower panel), the magnetic field values at which the $E_{x(y)}^0\leftrightarrow E_{x(y)}^{-1}$ ESLACs occur, depend only very weakly on the value of $g_l$ and we therefore consider these errors to be close to negligible. 

\begin{figure*}[h]
	\centering
	\includegraphics[width=1\textwidth]{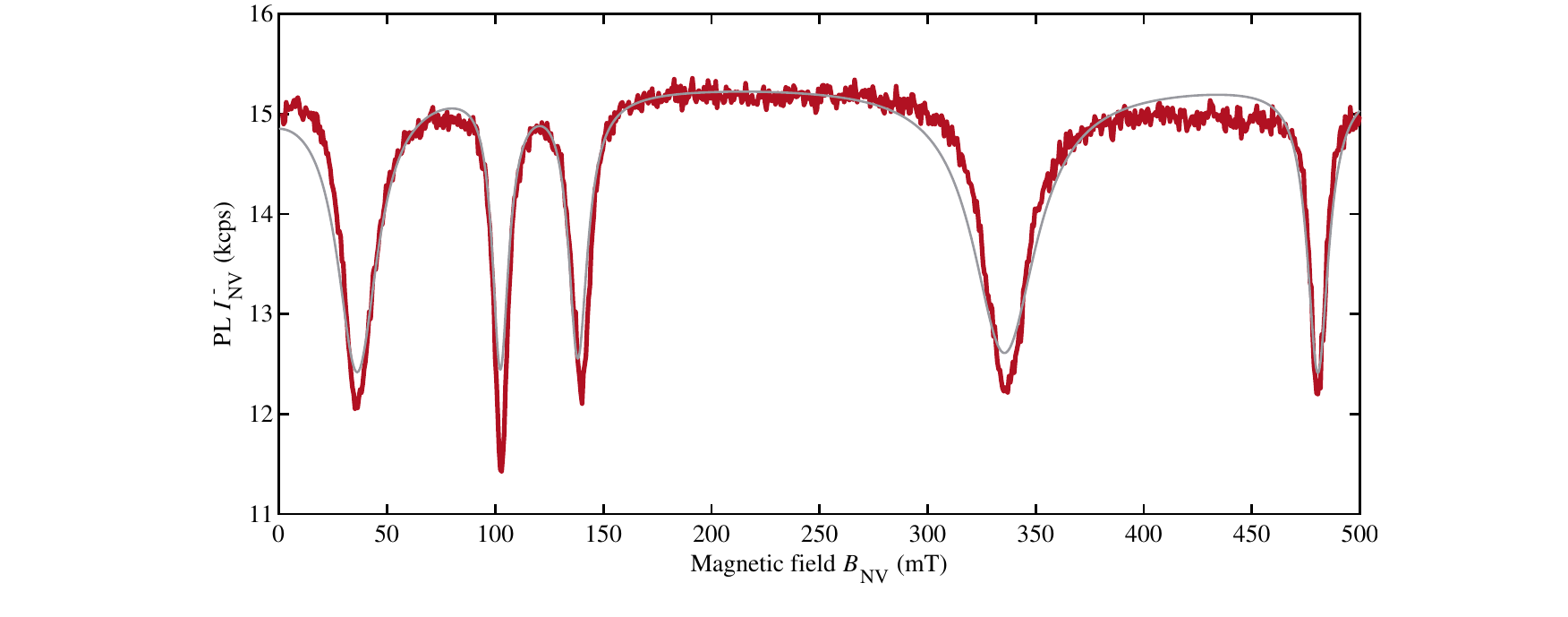}
	\caption{\NVm{} photoluminescence (PL) signal, $I_{\rm PL}^{-}$, as a function of magnetic field, $B_{\rm NV}$, for NV1-S1.
	The fitting values are:
	$\delta_\perp = 4.444\pm0.003GHz$,
	$\phi_{\delta} = -11.6\pm1.3^{\circ}$,
	$g_l = 0.1395\pm 0.0009$;
	B field misalignment:
	$\theta = 1.14\pm0.03^{\circ}$;
	Excitation and scaling parameters:
	$\beta_{E_x} = 0.0221\pm0.0003$,
	$\beta_{E_y} = 0.0242\pm0.0005$,
	$I^{-}_{\rm bck} = 4378\pm92$cps;
	}
	\label{fig:NV1_S1}
\end{figure*}

\begin{figure*}[h]
	\centering
	\includegraphics[width=1\textwidth]{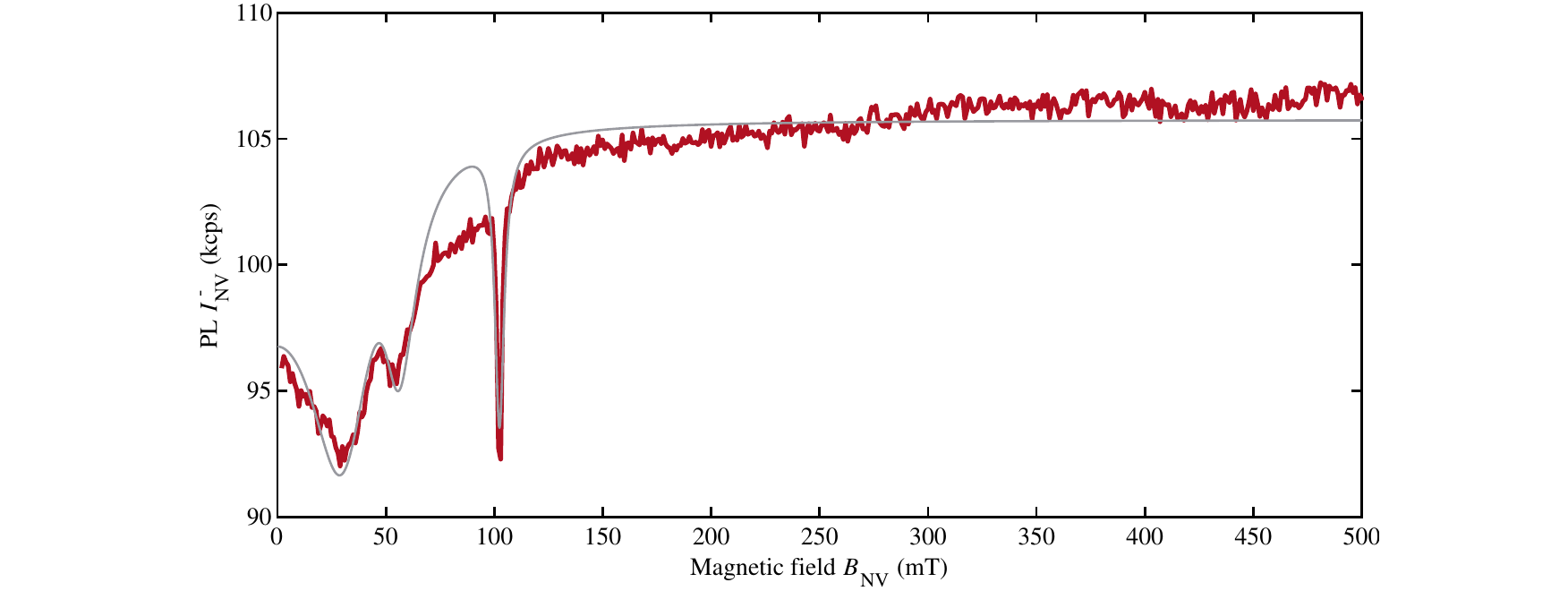}
	\caption{\NVm{} photoluminescence (PL) signal, $I_{\rm PL}^{-}$, as a function of magnetic field, $B_{\rm NV}$, for NV2-S2.
	The fitting parameters are:
	$\delta_\perp = 38.6\pm0.7$GHz,
	$\phi_{\delta} = 1\pm59^{\circ}$;
	B field misalignment:
	$\theta = 0.61\pm0.03^{\circ}$;
	Excitation and scaling parameters:
	$\beta_{E_x} = 0.085\pm0.005$,
	$\beta_{E_y} = 0.186\pm0.014$,
	$I^{-}_{\rm bck} = 70910.429\pm244.9818$cps;
	}
	\label{fig:NV2_S2}
\end{figure*}

\begin{figure*}[h]
	\centering
	\includegraphics[width=1\textwidth]{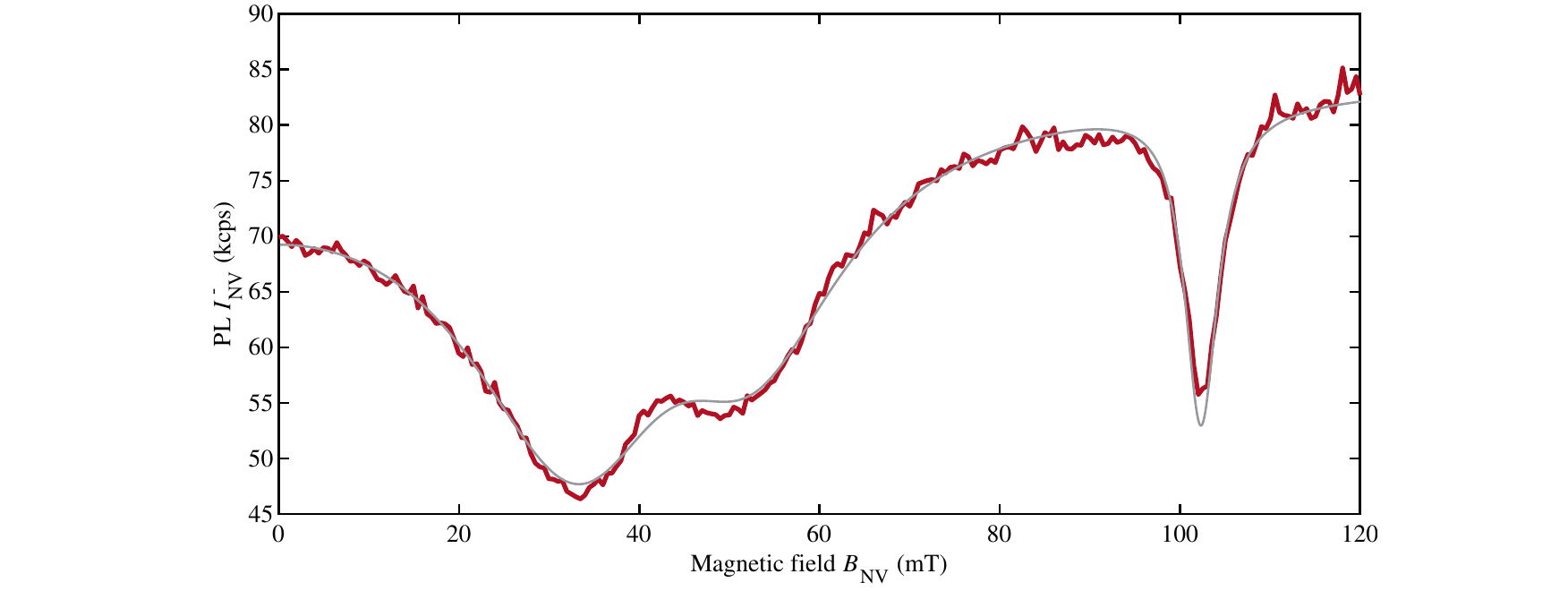}
	\caption{\NVm{} photoluminescence (PL) signal, $I_{\rm PL}^{-}$, as a function of magnetic field, $B_{\rm NV}$, for NV3-S2.
	The fitting parameters are:
	$\delta_\perp = 55.9\pm0.6$GHz,
	$\phi_{\delta} = -32.2\pm0.6^{\circ}$;
	B field misalignment:
	$\theta = 0.502\pm0.010^{\circ}$;
	Excitation and scaling parameters:
	$\beta_{E_x} = 0.240\pm0.006$,
	$\beta_{E_y} = 0.64\pm0.02$,,
	$I^{-}_{\rm bck} = 9091\pm231$cps;
	}
	\label{fig:NV3_S2}
\end{figure*}

\begin{figure*}[h]
	\centering
	\includegraphics[width=1\textwidth]{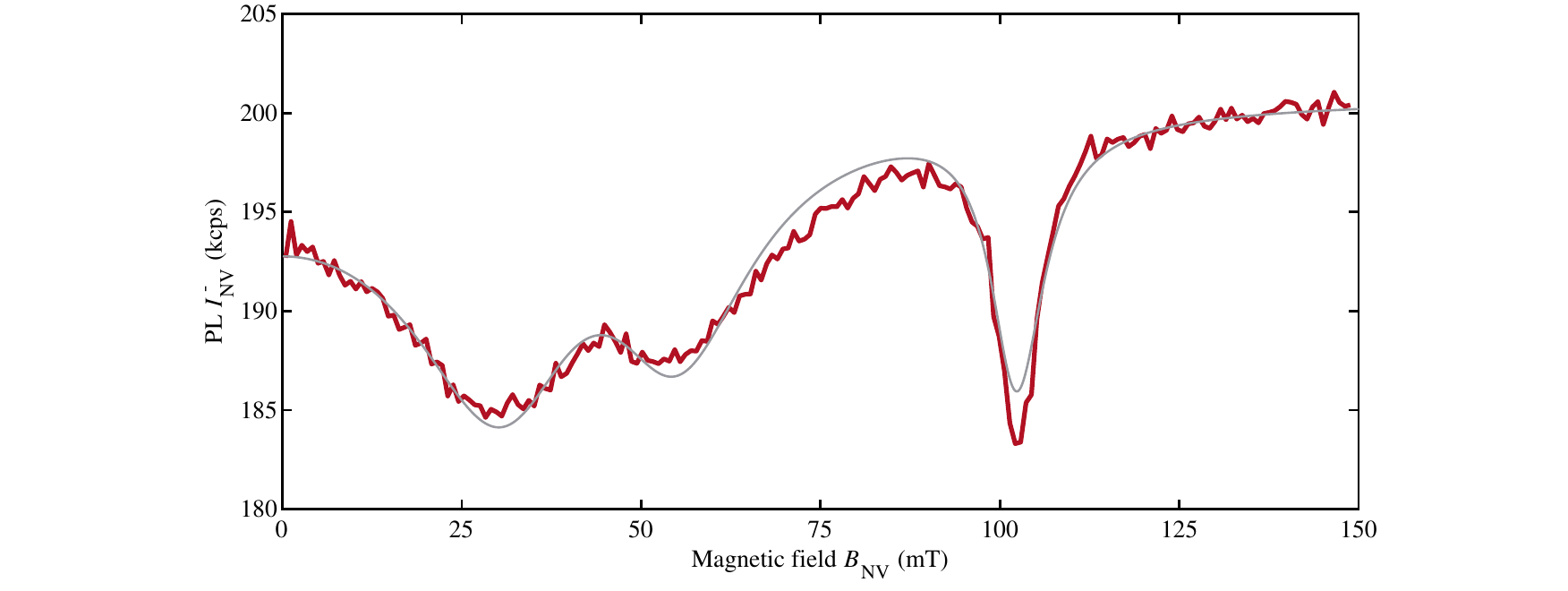}
	\caption{\NVm{} photoluminescence (PL) signal, $I_{\rm PL}^{-}$, as a function of magnetic field, $B_{\rm NV}$, for NV4-S3.
	The fitting parameters are:
	$\delta_\perp = 42.3\pm0.5$GHz,
	$\phi_{\delta} = -26\pm2^{\circ}$;
	B field misalignment:
	$\theta = 1.06\pm0.03^{\circ}$;
	Excitation and scaling parameters:
	$\beta_{E_x} = 0.102\pm0.003$,
	$\beta_{E_y} = 0.164\pm0.007$,
	$I^{-}_{\rm bck} = 158011\pm234$cps;
	}
	\label{fig:NV4_S3}
\end{figure*}

\begin{figure*}[h]
	\centering
	\includegraphics[width=1\textwidth]{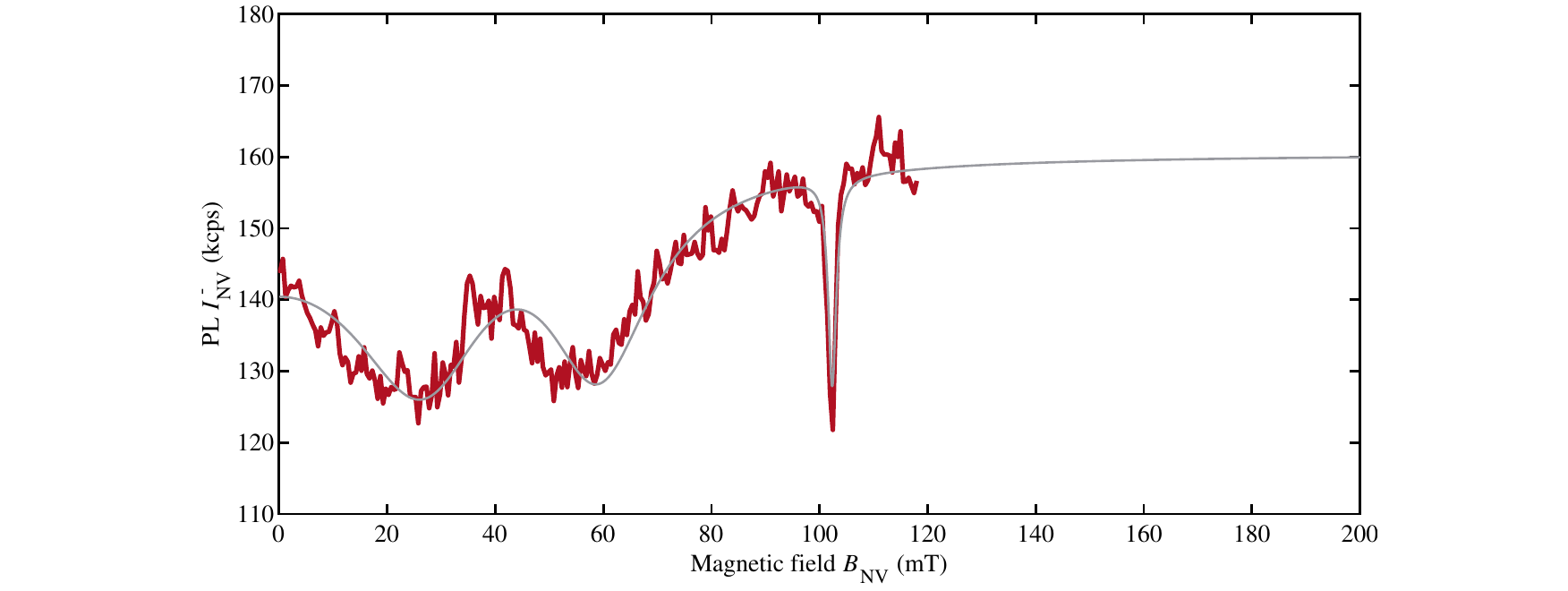}
	\caption{\NVm{} photoluminescence (PL) signal, $I_{\rm PL}^{-}$, as a function of magnetic field, $B_{\rm NV}$, for NV5-S2.
	The fitting parameters are:
	$\delta_\perp = 32.0\pm0.5$ GHz,
	$\phi_{\delta} = -29\pm3^{\circ}$;
	B field misalignment:
	$\theta = 0.20\pm0.02^{\circ}$;
	Excitation and scaling parameters:
	$\beta_{E_x} = 0.43\pm0.03$,
	$\beta_{E_y} = 0.52\pm0.05$,
	$I^{-}_{\rm bck} = 83198\pm883$ cps;
	The strain values are used to calculate the ESR contrast and the sensitivity in Fig.~4. The scaling parameters have been adapted due to the different laser power during the ESR experiment.
	}
	\label{fig:NV5_S2}
\end{figure*}

All of the fit results are summarized in Table.\,\ref{Table: Fit results}, and the fits from the NV's mentioned in the main text are shown in Figures\,\ref{fig:NV1_S1} to \ref{fig:NV5_S2}.

\begin{table}\label{Table: Fit results}
	\begin{tabular}{l l l l l l l l }
		\hline
		NV-Sample \qquad & $\delta_\perp$ (GHz)  \qquad & $\phi_\delta$ ($^\circ$)  & $g_l$  \qquad\qquad\qquad & $\theta_B$ ($^\circ$) & $\beta_{E_x}$ \qquad\quad\, & $\beta_{E_y}$  \qquad\quad\, & $I_{\rm bck}^{-}$ (cps) \\
		\hline
		NV1-S1 & 4.444(3) & -11.6(1.3) & 0.1395(9) & 1.14(3) & 0.0221(3) & 0.0242(5) & 4378(92) \\
		NV2-S2 & 38.6(7) & 1(59) & (fixed to 0.14) & 0.61(3) &  0.085(5) & 0.186(14) &  70910(244) \\
		NV3-S2 & 55.9(6) & -32.2(6) & (fixed to 0.14) &  0.502(10) & 0.240(6) & 0.64(2) & 9091(231) \\
		NV5-S2 & 32.0(5) & -29(3) & (fixed to 0.14) & 0.20(2) & 0.43(3) &  0.52(5) & 83198(883) \\
		NV4-S3 & 42.3(5) & -26(2) & (fixed to 0.14) &  1.06(3) &  0.102(3) & 0.164(7) & 158011(234) 
	\end{tabular}
\caption{Summary of the fit results of the $I_{\rm PL}^-$ spectra.}
\end{table}

\subsection{Magnetic field dependent photoluminescence of the two NV charge states}\label{Sec: FittingNV}

The \NVz{} PL can be fitted with the above presented 12-level model including the \NVz{} ground and excited states.
The physical parameters for the NV ($\theta_B$, $\delta_\perp$, $\phi_\delta$, $\beta_{E_{x}}$, $\beta_{E_{y}}$) which have been extracted from the fit to the corresponding \NVm{} PL data, are used as constant parameters in the model and are not fitted.
As a pumping parameter for the \NVz{} and the laser power dependent shelf-ionization rate from the singlet we take the average between the two pumping parameters to the two \NVm{} ES orbitals. 

The only free parameters left are scaling parameters such that:
\begin{equation}
	I^{0}_{\rm PL} = f(\eta_{NV^{0}}, I^{0}_{\rm bck})
\end{equation}
where $\eta_{NV^0}$ is the collection efficiency for the \NVz{} PL and $I^{0}_{\rm bck}$ are the background counts in the wavelength spectrum of the collected \NVz{} PL.

Due to the new decay channels in the model also the \NVm{} PL needs to be rescaled when using the 12-level model.
We found that merely with leaving the $I^{-}_{\rm PL}$ as a free parameter gives good accordance with the data.
This leaves the fit to the \NVm{} PL with:
\begin{equation}
	I^{-}_{\rm PL} = f(I^{-}_{\rm bck})
\end{equation}
where $I^{-}_{\rm bck}$ is the background counts in the wavelength spectrum of the collected \NVm{} PL.

The \NVz{} PL exactly mirrors the dips of the \NVm{} with peaks. This behavior cannot be explained solely by the known ionization and recombination processes between the two charge states. Considering only them would lead to the same dips  as there are in the \NVm{} PL instead of the observed peaks. However when introducing the recently proposed channel from the \NVm{} singlet to the \NVz{} ground state this behavior can be explained with our model.
Even when leaving more free fit parameters, without a certain minimal value for the shelf-ionization rate the behavior of the data cannot be reproduced. The fits produce a larger value of background fluorescence counts in the fitting values of the \NVm{} data. This increase could indicate some small discrepancy with the ionization, recombination and shelf-ionization rates and/or their dependence on laser power.

\begin{table}\label{Table: Fit results for 12-level model with fit parameter taken from the 10-level model fit}
	\begin{tabular}{l l l l l l l l }
		\hline
		NV-Sample \qquad & $\eta_{NV^{0}}$ \qquad\qquad\qquad & $I^{0}_{\rm bck}$ (cps) \qquad\qquad& $I^{-}_{\rm bck}$ (cps) \\
		\hline
		NV1-S1 & 0.00132(3) & 1258(11) & 7980(8) \\
		NV2-S2 & 0.001820(7) & 7963(10) & 79176(52)
	\end{tabular}
\caption{Summary of the fit results of the $I_{\rm PL}^0$ and  $I_{\rm PL}^-$ spectra.}
\end{table}

\begin{figure*}[th]
	\centering
	\includegraphics[width=1\textwidth]{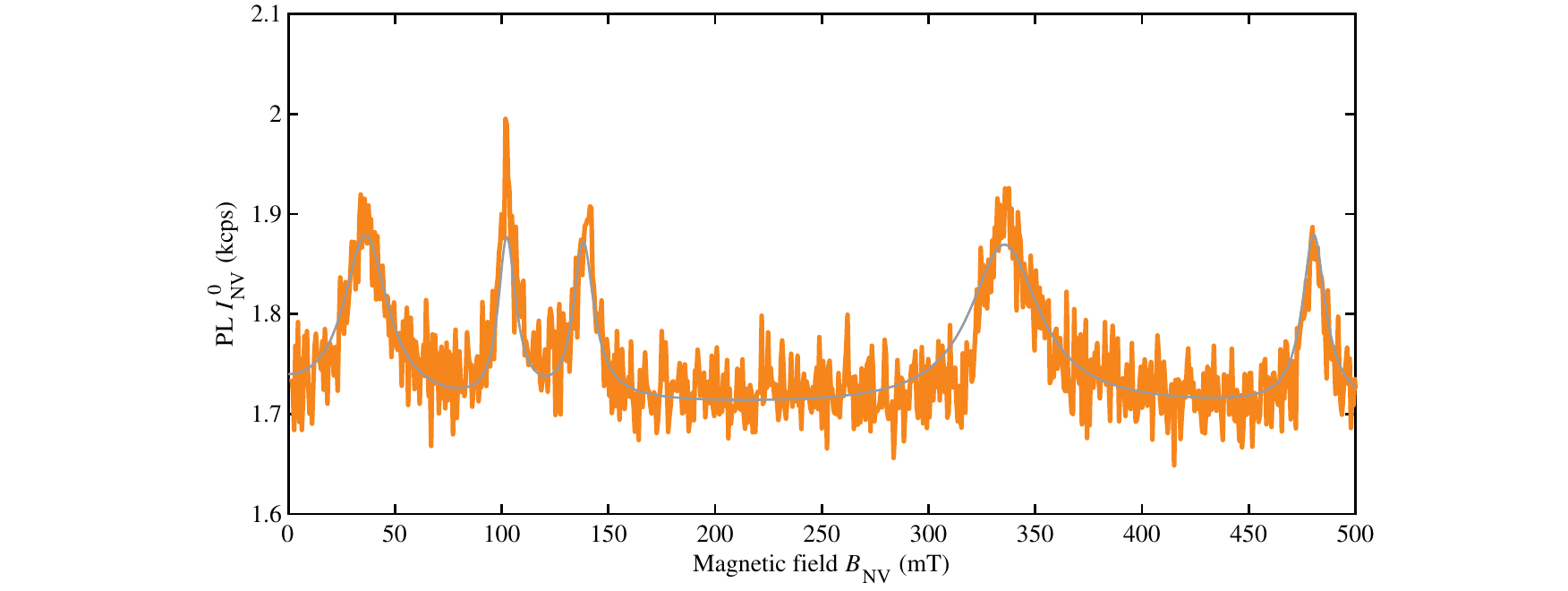}
	\caption{\NVm{} photoluminescence (PL) signal, $I_{\rm PL}^{0}$, as a function of magnetic field, $B_{\rm NV}$, for NV1-S1.
	The fitting values for \NVz{} and \NVm{} PL are:
	$\eta_{NV^{0}}=0.00132\pm0.00003$ and the background fluorescence for \NVz{}
	$I^{0}_{\rm bck}=1258\pm11$;
	the background fluorescence for \NVm{}
	$I^{-}_{\rm bck}=7980\pm8$;
	}
	\label{fig:NV1_S1_NV0}
\end{figure*}

\begin{figure*}[th]
	\centering
	\includegraphics[width=1\textwidth]{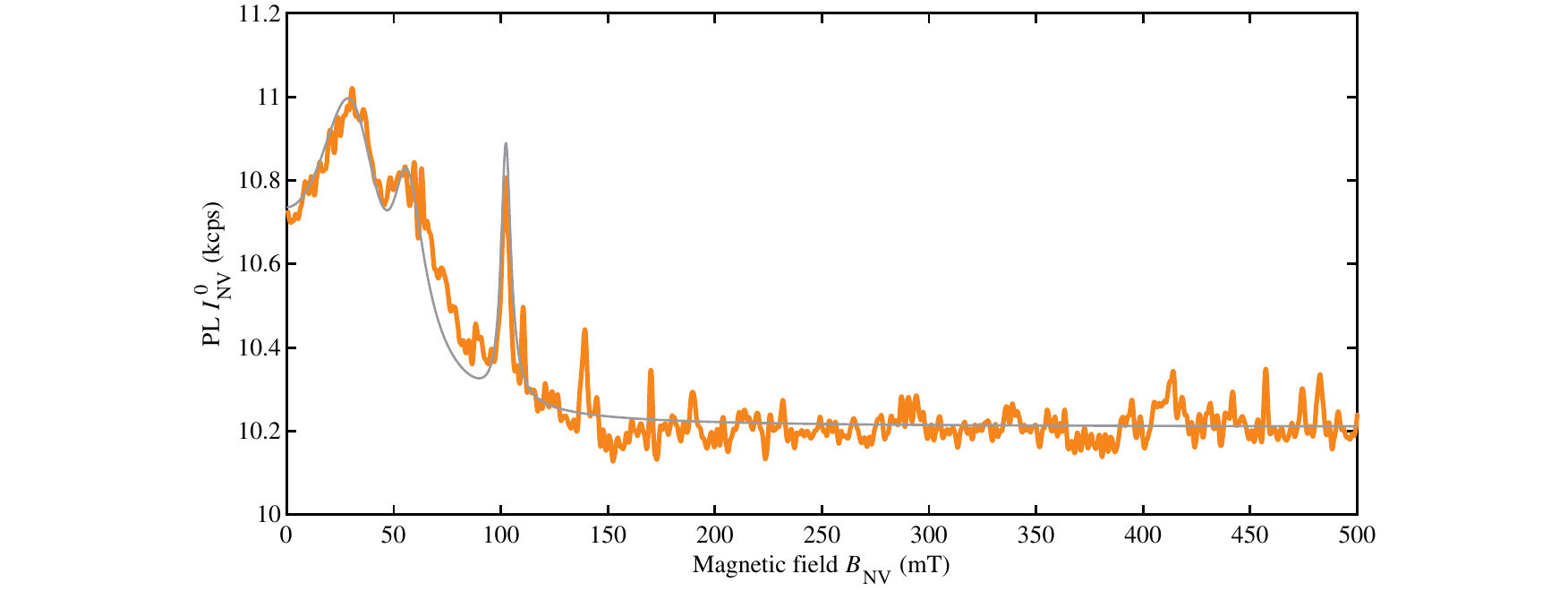}
	\caption{\NVm{} photoluminescence (PL) signal, $I_{\rm PL}^{0}$, as a function of magnetic field, $B_{\rm NV}$, for NV2-S2.
	The fitting values for \NVz{} and \NVm{} PL are:
	$\eta_{NV^{0}}= 0.001820\pm0.000007$ and the background fluorescence for \NVz{}
	$I^{0}_{\rm bck}=7963\pm10$;
	 the background fluorescence for \NVm{}
	$I^{-}_{\rm bck}=79176\pm52$;
	}
	\label{fig:NV2_S2_NV0}
\end{figure*}

\newpage
\section{Experimental details}
\subsection{Experimental apparatus}\label{Sec: Set up}

\begin{figure*}[t]
	\centering
	\includegraphics[width=1\textwidth]{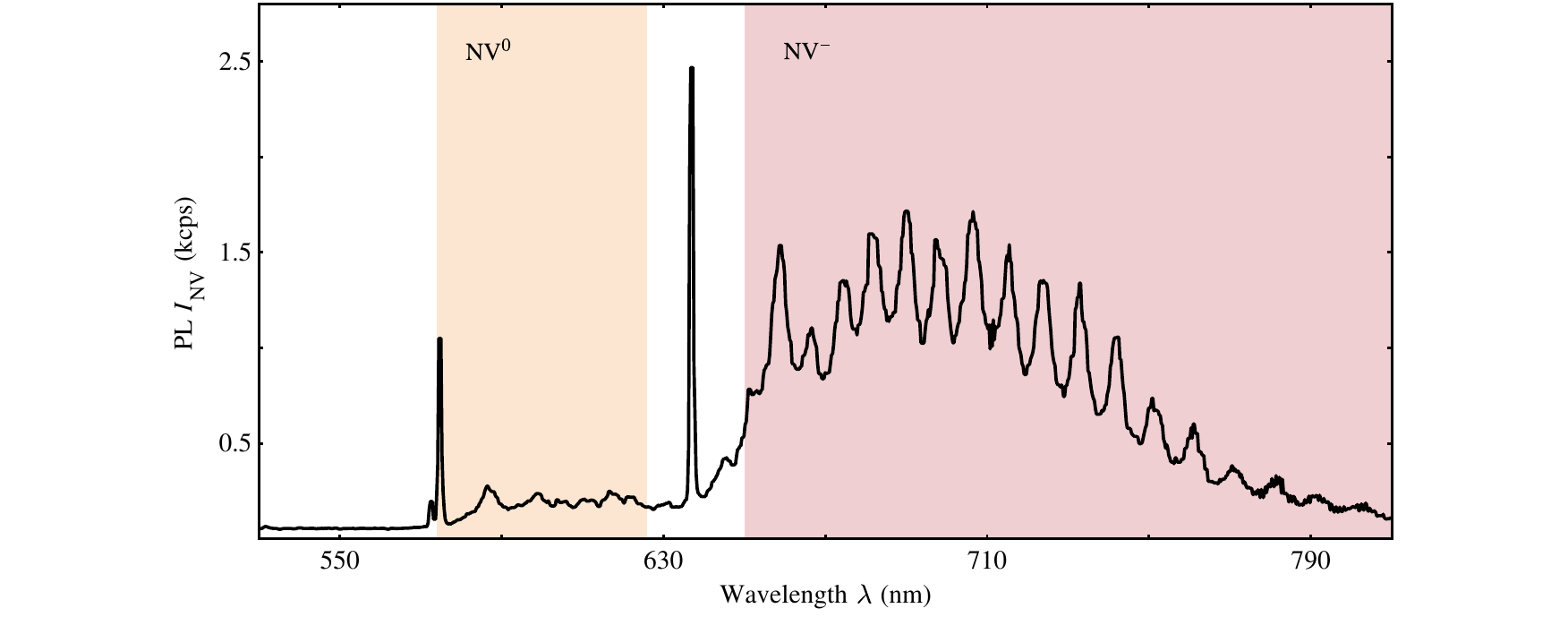}
	\caption{
		Optical emission spectrum of a single NV in a diamond nanostructure\cite{Hedrich2020a,neuPhotonicNanostructures1112014}. The highlighted regions indicate the regions of collected PL to separate the slightly overlapping spectra of the charge states \NVz{} (orange) and \NVm{}(red).
	}
	\label{fig:Spectrum}
\end{figure*}

The NV experiments are performed in a closed-cycle refrigerator (Leiden Cryogenics, CF-CS81) with a $(B_x, B_y, B_z) = (1,1,5)$~\si{\tesla} vector magnet.
All of the measurements were all performed at a temperature of around \SI{4}{\kelvin}.
The cryostat possesses free-space optical access allowing the optical initialization and read-out of the NV through a confocal microscope with a green excitation laser at \SI{532}{\nano\meter} (Coherent Compass 315M), where the excitation polarization is controlled with a liquid crystal polarization rotator (Thorlabs LCR-532).
The NV PL read-out is performed with an avalanche photo-diode (Excelitas SPCM-AQRH-33) with a \SI{650}{\nano\meter} long-pass filter (Thorlabs FELH0650) for the NV$^{-}$ or a \SI{600/52}{\nano\meter} bandpass filter (Semrock  FF01-600/52-25) for the NV$^{0}$.
An example of these filters on a diamond nanopillar can be seen in Figure\,\ref{fig:Spectrum}.
The NV spectrum is obtained with a spectrograph (Princeton Instruments, HRS-500) and imaged with a low-noise camera (PIXIS 100).

The microwave control of the NV spin was performed using a constant microwave driving field from a signal generator (SRS SG384) that was modulated via an IQ-modulator (Polyphase microwave, AM0350A Quadrature Modulator) and an AWG (Spectrum generator NETBOX DN2.663-04). The microwave are then applied to the NV spin via a 25~$\mu$m thick Al wire that the NV is brought into close proximity with. The NV photoluminescence is then measured via gating the output of the APD with a TTL-router (in-house built, SP995) which is controlled via the AWG and whose output is measured with a DAQ (NI PXIE-1073, PXI-6220). 

\subsection{Samples details}\label{Sec: Samples}

In the following, we provide additional details on the samples used in this work:

{\bf Sample S1} is identical to the sample used in ref.\,\cite{Robledo2011a}, where extensive details on the sample and sample fabrication are given. The sample is a $(100)-$oriented, chemical vapor deposition (CVD) grown type IIa ``electronic grade'' diamond from Element 6. Solid immersion lenses were fabricated by focused ion beam milling around pre-localized, naturally occurring NV centers located $5-15~\mu$m below the sample surface.

{\bf Sample S2} consisted of a commercially available, $(100)-$oriented, CVD grown type IIa ``electronic grade'' diamond from Element 6. NV centers were created by $^{14}$N$^+$ ion implantation at $12~$keV and subsequent annealing as described elsewhere\,\cite{Appel2016}. This procedure results in NV centers located $\sim10~$nm from the diamond surface. To increase photon collection efficiency from the NVs, cylindrical diamond nanopillars of $\sim200~$nm diameter were fabricated in the sample surface\,\cite{Appel2016}. The sample fabrication procedure was tailored to yield an average of one NV center per diamond pillar. 

{\bf Sample S3}: is identical to the sample used in ref.\,\cite{Neu2014a}, where extensive details on the sample and sample fabrication are given. The sample was CVD grown by the group of J. Achard and A. Tallaire with methods described in detail in Ref.\,\cite{Tallaire2014}. NV centers were created in growth through controlled incorporation of N gas into the growth reactor. Under the growth conditions used here, these leads to NV centers whose quantization axis is preferentially aligned with the 111 growth direction and which typically show excellent spin properties\,\cite{Lesik2014}. 
We note that the sample exhibits elevated fluorescence background levels from SiV centers which were inadvertedly introduced during sample growth.

Prior to our experiments, samples S2 and S3 were acid cleaned using a well-established acid cleaning technique, details in Ref.\,\cite{Brown2019}. This sample cleaning method leaves the diamond surface predominantly O-terminated. No acid cleaning was employed for sample S1 to preserve pre-existing antenna structures on the sample surface. Given the micron-scale depth of the NVs examined in sample S1, sample cleaning and surface termination are or minor relevance to the behavior of NVs studies in this sample. 


\subsection{$g_l$ factor measurements}

\begin{figure*}[th]
	\centering
	\includegraphics[width=1\textwidth]{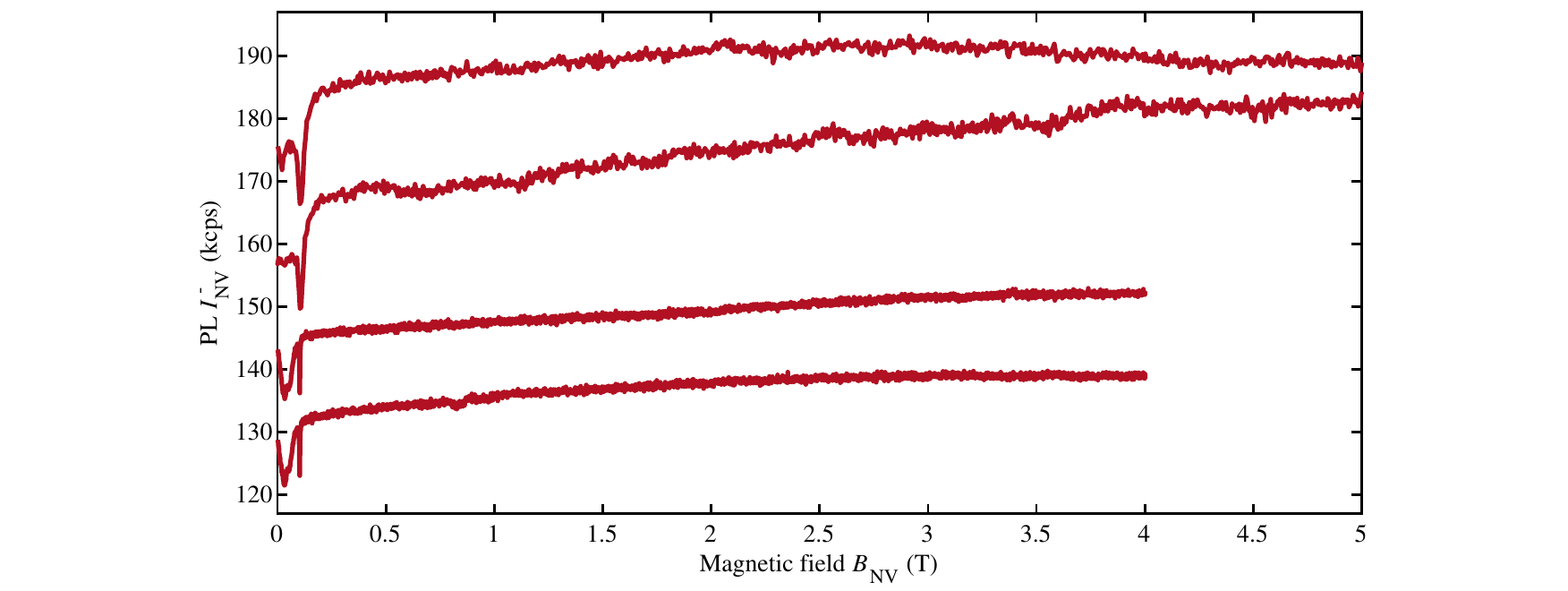}
	\caption{\NVm{} photoluminescence (PL) signal, $I_{\rm PL}^{-}$ as a function of magnetic field, $B_{\rm NV}$, for multiple NVs showed with an offset.
	According to our model all of the shown NV PL spectra should show PL dips in the shown magnetic field range.
	They are exemplary and show full field scans.
	Most of the performed scans were regions of magnetic field investigating of various parts of the magnetic field.
	To determine the strain appropriately we performed high resolution scans at low magnetic fields.
	}
	\label{fig:gfactor}
\end{figure*}

For the measurement of the inter-branch level anti crossings $E_{x(y)}\leftrightarrow E_{y(x)}$ at high magnetic fields, the $111$ oriented sample S3 was introduced to maximize the magnitude of $B_{\rm NV}$ that we could access in our experimental setup equipped with a 1T/1T/5T vector magnet.
With all NVs, we proceeded as described for NV4-S3 in the main text: We first characterized NVs with PL versus B measurements al ``low'' magnetic fields (c.f. Fig.2a in the main text) to determine the effective strain parameter.
After that we investigated $I_{\rm PL}^{-}(B_{\rm NV})$ for $B_{\rm NV}$ up to $5$~T.
Due to slow magnetic field ramp speeds and occasional signal drifts, we mostly performed magnetic field sweeps over smaller intervals in $B_{\rm NV}$, and only few measurements cover the entire range -- those data are shown in Fig.\,\ref{fig:gfactor}. 
Some of the data include a low-frequency envelope, which results from system drifts during the magnetic field sweeps. 
The NVs we investigated in S3 all showed $\delta_\perp$-values on the range $30...50~$GHz.
According to our model, these NVs should all exhibit their $E_{x(y)}\leftrightarrow E_{y(x)}$ inter-branch ESLACs, and a corresponding $I_{\rm PL}^{-}$ drop, in the centre of the magnetic field sweep range shown in Fig.\,\ref{fig:gfactor}, unless $g_l$ is significantly enhances, as discussed in the main text.

\subsection{Magnetic field sensitivity}

In order to probe the effect of the excited state level anticrossings on the performance of magnetic field sensing, single NV spins in nanopillars (Sample S2) were brought into close proximity (within 100~$\mu$m) of a 25~$\mu$m thick Al wire.  
To remove frequency dependent performance of the microwave circuitry, at each magnetic field position RABI measurements were performed to tune the microwave driving power to maintain an average $\pi$-time of 200~ns (power broadened, results shown in main script) and 1~$\mu$s (hyperfine resolved).
The optically detected magnetic field measurements where then performed with a standard pulsed ODMR technique\,\cite{Dreau2011} with these driving times and fitted with Lorentzian functions.
 
We were not able to detect any statistically difference in the variation in sensitivity between the power broadened and hyperfine resolved ODMR.
Additionally, we did not observe any additional effect on the nuclear spin polarization, which is starkly different to the room temperature excited state level anticrossing, where polarization of greater than 90\% have been observed\,\cite{Jacques2009}.
  
To compare the sensitivity and contrast with the model, we first fit the \NVm{} $I_{\rm PL}$ spectra to obtain the parameters from the model.
These fit parameters are then used to predict the contrast of the ODMR as a function of magnetic field, $\mathcal{C}(B)$. 
The combination of the fitted $I_{\rm PL}^0(B)$ and the predicted $\mathcal{C}(B)$ are then used in conjunction with the average ODMR width (small variation due to rabi tuning) to predict the sensitivity, $\eta(B)$.

\subsection{Polarization dependence of the NV PL}\label{Sec: Polarization}

The optically pumping to the excited state orbital doublet has a polarization dependence.
Therefore it is possible to control the prominence of the dips in the PL which arise due to level anticrossing.
The control is limited due to the non-radiative relaxation after pumping\,\cite{Fu2009a}.

Here we controlled the rotation of the 532nm excitation via a liquid crystal rotator with a 1000:1 extinction ratio which was placed after the dichroic mirror before the objective.

\begin{figure*}[th]
	\centering
	\includegraphics[width=1\textwidth]{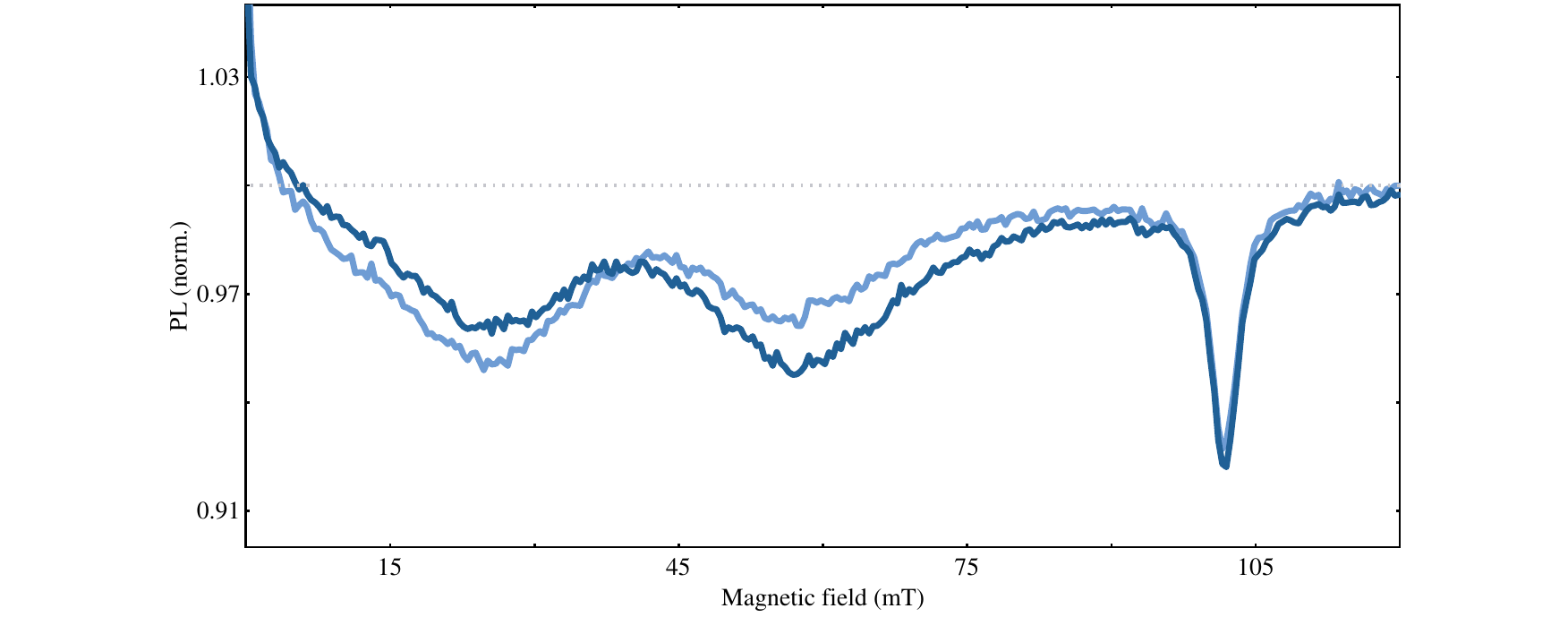}
	\caption{Normalized \NVm{} photoluminescence (PL) signal, $I_{\rm PL}^{-}$, as a function of magnetic field, $B_{\rm NV}$, for two orthogonal laser polarizations of an NV in S3. The two inter-branch LAC are well resolved. The initial drop in counts is attributed  to a nearby NV which is quenched with magnetic field. 
	}
	\label{fig:S3_Polarization}
\end{figure*}

\begin{figure*}[h]
	\centering
	\includegraphics[width=1\textwidth]{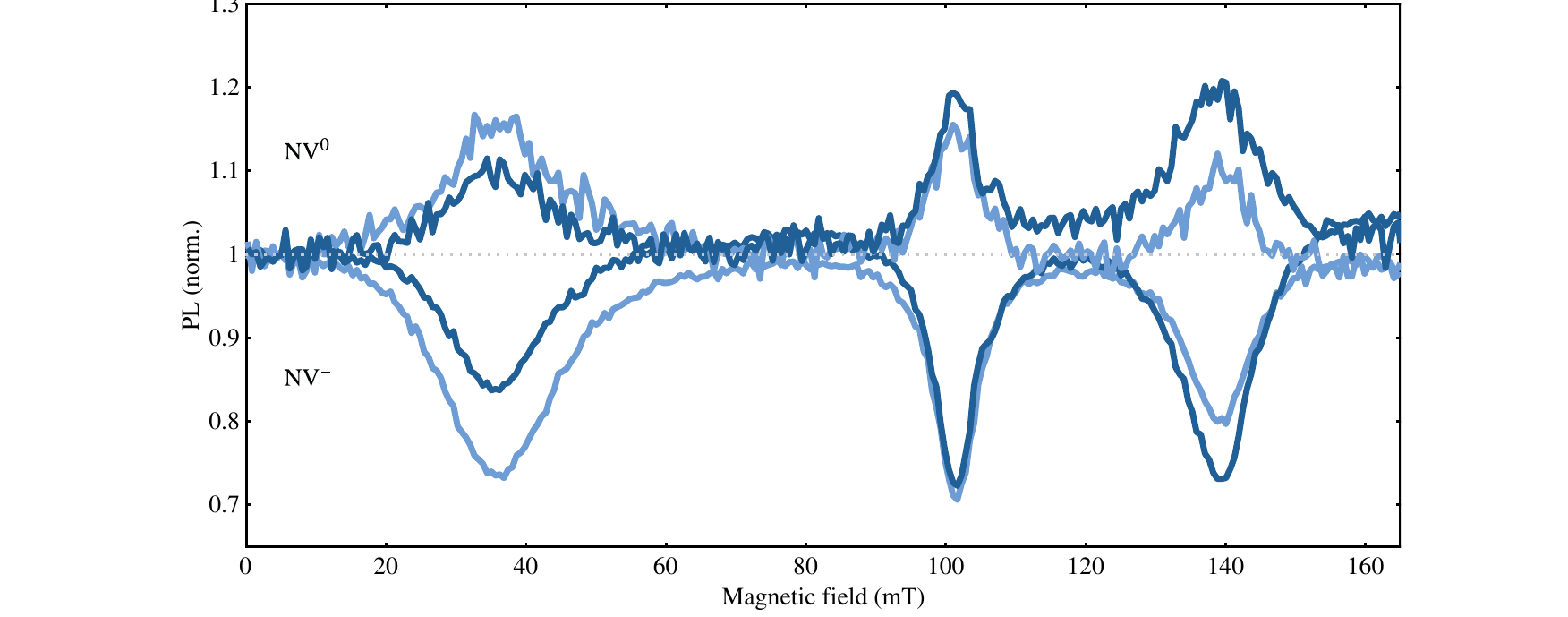}
	\caption{Normalized \NVm{} photoluminescence (PL) signal, $I_{\rm PL}^{-}$, and normalized  \NVz{} photoluminescence (PL) signal, $I_{\rm PL}^{0}$, as a function of magnetic field, $B_{\rm NV}$, for the low-strain NV1-S1. Two orthogonal laser polarizations have been used. The \NVz{} signal shows a dependence on laser polarization at the same positions as the \NVm{} does. When the \NVm{} countrate decreases at an ES LAC, the corresponding \NVz{} countrate increases and vice versa.
	}
	\label{fig:NV1_S1_Polarization}
\end{figure*}

In both the higher (Fig.\,\ref{fig:S3_Polarization}) and low stress regime (Fig.\,\ref{fig:NV1_S1_Polarization}) we found that we could change the contrast of the peaks.
While this is not a significant amount of tuning, it is possible that further tuning is possible through changing the wavelength of the excitation as well as further optimization of the polarization control.

The \NVz{} PL also shows a dependence on the laser polarization which can be seen in the upper part of Fig.\,\ref{fig:NV1_S1_Polarization}.

\end{document}